\documentclass[debug, preprint, twocolumn]{rmaa}
\usepackage{natbib}
\usepackage{graphicx} 
\usepackage{amsmath}
\usepackage{hyperref}
    
\usepackage{paralist}

\usepackage{psfrag,color}

\usepackage[latin1]{inputenc}

{\catcode`\$=\active
}%

\newcommand\Moon{\astro{\char10}}
\newcommand\Aries{\astro{\char11}}%
\newcommand\Libra{\astro{\char17}}%
\newcommand{\Cloudy}{\textsc{Cloudy}}
\newcommand{\Hazy}{\textsc{Hazy}}
\newcommand{\cdCommand}[1]{\textbf{#1}}

\newcommand{\cdFilename}[1]{\texttt{#1}}
\newcommand{\cdRoutine}[1]{\emph{#1}}

\font\manual=manfnt at 7pt \def\dbend{\hbox{\raise0.9ex\hbox{\manual\char127\hspace{0.6em}}}}

\newcommand\OR{\texorpdfstring{\ensuremath{\vert}}{|}}

\providecommand{\e}[1]{\ensuremath{\times 10^{#1}}}

\newcommand\Ion[2]{\ensuremath{\mathrm{#1\,\scriptstyle #2}}}

\newcounter{INTERNALionstage}
\providecommand{\ion}[2]{
  \setcounter{INTERNALionstage}{#2}%
  \Ion{#1}{\Roman{INTERNALionstage}}}

\def\gtsim{\mathrel{\hbox{\rlap{\hbox{\lower4pt\hbox{$\sim$}}}\hbox{$>$}}}}
\def\lesssim{\mathrel{\hbox{\rlap{\hbox{\lower4pt\hbox{$\sim$}}}\hbox{$<$}}}}

\def\A{{\rm\thinspace \AA}}

\def\g{{\rm\thinspace g}}

\def\K{{\rm\thinspace K}}

\def\m{{\rm\thinspace m}}

\def\micron{\hbox{$\mu$m}}

\def\ps{{\rm\thinspace s^{-1}}}
\def\pcc{{\rm\thinspace cm^{-3}}}

\def\s{{\rm\thinspace s}}

\def\W{{\rm\thinspace W}}

\def\ps{\mbox{$\s^{-1}\,$}}

\def\hi{\mbox{{\rm H~{\sc i}}}}

\def\feii{\mbox{{\rm Fe~{\sc ii}}}}

\def\htwo{\mbox{{\rm H}$_2$}}

\def\h0{\mbox{{\rm H}$^0$}}
\DeclareMathAlphabet{\vib}{OML}{cmm}{m}{it}

\usepackage{common/aas_macros}

\title{The 2023 release of \Cloudy}
\author{Marios Chatzikos\altaffilmark{1}, 
Stefano Bianchi\altaffilmark{2},
Francesco Camilloni\altaffilmark{3},
Priyanka Chakraborty\altaffilmark{4},
Chamani M. Gunasekera\altaffilmark{1},
Francisco Guzm\'an\altaffilmark{1,5},
Jonathan S. Milby\altaffilmark{6},
Arnab Sarkar\altaffilmark{7},
Gargi Shaw\altaffilmark{8},
Peter A. M. van Hoof\altaffilmark{9},
Gary J. Ferland\altaffilmark{1}}

\altaffiltext{1}{Physics \& Astronomy, 
University of Kentucky, Lexington, Kentucky, USA}
\altaffiltext{2}{Dipartimento di Matematica e Fisica, Universit\`a degli Studi Roma Tre, via della Vasca Navale 84, 00146 Roma, Italy}
\altaffiltext{3}{Max-Planck-Institut f\"ur extraterrestrische Physik, Giessenbachstra{\ss}e, 85748 Garching, Germany}
\altaffiltext{4}{Center for Astrophysics $\vert$ Harvard \& Smithsonian, Cambridge,
MA, USA}
\altaffiltext{5}{Now at Physics \& Astronomy, 
University of North Georgia, Dahlonega, Georgia, USA}
\altaffiltext{6}{Arts \& Sciences Hive, 
University of Kentucky, Lexington, Kentucky, USA}
\altaffiltext{7}{Kavli Institute for Astrophysics and Space Research,
Massachusetts Institute of Technology, 70 Vassar St, Cambridge, MA, USA}
\altaffiltext{8}{Department of Astronomy and Astrophysics, Tata Institute of Fundamental Research, Mumbai 400005, India}
\altaffiltext{9}{Royal Observatory of Belgium, Ringlaan 3, 1180 Brussels, Belgium}

\fulladdresses{
 \item S.{} Bianchi, Dipartimento di Matematica e Fisica, Universit\`a degli Studi Roma Tre, via della Vasca Navale 84, 00146 Roma, Italy (stefano.bianchi@uniroma3.it).
  \item F.{} Camilloni, Max-Planck-Institut f\"ur extraterrestrische Physik, Giessenbachstra{\ss}e, 85748 Garching, Germany (fcam@mpe.mpg.de).
  \item P.{} Chakraborty, Center for Astrophysics $\vert$ Harvard \& Smithsonian, Cambridge, MA (priyanka.chakraborty@cfa.harvard.edu).
 \item M.{} Chatzikos, Physics \& Astronomy, University of Kentucky, Lexington KY 40506 (mchatzikos@uky.edu).
 \item G.{} Ferland, Physics \& Astronomy, University of Kentucky, Lexington KY 40506 (gary@uky.edu).
 \item C.{} Gunasekera, Physics \& Astronomy, University of Kentucky, Lexington KY 40506 (cmgunasekera@uky.edu).
 \item F.{} Guzm\'an, Physics and Astronomy, University of North Georgia, Dahlonega GA 30597 (francisco.guzmanfulgencio@ung.edu).
 \item J.{} Milby, Arts \& Sciences Hive, University of Kentucky, Lexington KY 40506 (jon@uky.edu).
 \item A.{} Sarkar, Kavli Institute for Astrophysics and Space Research,
Massachusetts Institute of Technology, 70 Vassar St, Cambridge, MA 02139, USA (arnabsar@mit.edu)
 \item G.{} Shaw, Department of Astronomy and Astrophysics, Tata Institute of Fundamental Research,  
Mumbai 400005, India (gargishaw@gmail.com).
 \item P.{} A.{} M.{} van Hoof, Royal Observatory of Belgium, Ringlaan 3, 1180 Brussels, Belgium (p.vanhoof@oma.be).
 }

\shortauthor{Chatzikos et al}
\shorttitle{Cloudy's 2023 Release}

\listofauthors{M. Chatzikos et al}
\indexauthor{Chatzikos, M.}
\indexauthor{Ferland, G.J.}
\indexauthor{Guzm\'an, F.}
\indexauthor{Jonathan S. Milby}
\indexauthor{Shaw, G.}
\indexauthor{Sarkar, Arnab}
\indexauthor{Gunasekera, C.M.}
\indexauthor{Chakraborty, Priyanka}

\abstract{We describe the 2023 release of the spectral synthesis code \Cloudy.
Since the previous major release, migrations of our online services motivated
us to adopt {\tt git} as our version control system.
This change alone led us to adopt an annual release scheme, accompanied by a
short release paper, the present being the inaugural.
Significant changes to our atomic and molecular data have improved the accuracy
of \Cloudy{} predictions:
we have upgraded our instance of the Chianti database from version 7 to 10;
our H- and He-like collisional rates to improved theoretical values; 
our molecular data to the most recent LAMDA database,
and several chemical reaction rates to their most recent UDfA and KiDA values.
Finally, we describe our progress on upgrading \Cloudy{}'s capabilities
to meet the requirements of the X-ray microcalorimeters aboard the upcoming
{\it XRISM} and {\it Athena} missions, and outline future development that will
make \Cloudy{} of use to the X-ray community.}
\resumen{
Se describe el lanzamiento de la version 2023 del c\'odigo de s\'intesis espectral \Cloudy.
La migraci\'on de nuestros servicios online motiv\'o la adopci\'on de {\tt git}
como nuevo sistema de control de versiones.
Este cambio condujo a un plan de lanzamientos anuales, acompa\~nados de un art\'iculo breve, comenzando por el 
presente.
Cambios significativos en los datos at\'omicos y moleculares 
mejoran la exactitud de las predicciones de \Cloudy{} mediante las 
actualizaciones de la base de datos de Chianti de la version 
7 a la 10, las transiciones colisionales en iones de uno 
y dos electrones, los datos moleculares a la
version m\'as reciente de la base de datos LAMBDA y 
varias constantes de reacci\'on moleculares  
a los valores de UDfA y KiDA m\'as recientes.
Finalmente, se describe el proceso de adaptaci\'on 
de \Cloudy{} a los requisitos de los microcalor\'imetros a bordo de las
misiones {\it XRISM} y {\it Athena} y el progreso para hacer \
\Cloudy{} \'util para la comunidad de astrof\'isica de rayos X.   
}

\addkeyword{Atomic data}
\addkeyword{Astronomy software}
\addkeyword{Active galaxies}
\addkeyword{Computational methods}
\addkeyword{Galaxy clusters}
\addkeyword{Molecular data}

\begin{document}

\maketitle

\section{Introduction}

\Cloudy{} is an {\it ab initio} spectral synthesis code for
astrophysical plasmas ranging from far from equilibrium
to Local Thermodynamic Equilibrium (LTE) and Strict
Thermodynamic Equilibrium (STE).
Development started in 1978, and has been ongoing since then,
with each new release extending the physical systems the code
can model.
Previous review papers capture the state of the code at that time,
namely,
\citet{1998PASP..110..761F},
\citet{2013RMxAA..49..137F}
and
\citet{2017RMxAA..53..385F}.
Much of the physics is discussed in \citet{AGN3}.

In the past, we aimed to release code only following major
changes to the source code and our quantum physics data.
The underlying principle had been to deliver our users with
a product that would have maximal impact on their research.
A consequence of this policy had been infrequent releases,
with only seven (7) taking place in the period 1998-2017.
The 2006-2010 releases (C06, C07, C08, and C10) were not
accompanied by a review paper, which may have left some users
wondering how significant the changes in each new version were.
Subsequent releases were accompanied by a major review article of
\Cloudy's capabilities, which further delayed each
release.
In this dilemma, we had contemplated if a
better release policy might be pursued.

The changes described in Section~\ref{sec:access}
were the impetus we needed to adopt a new release policy.
This has essentially been enabled by transitioning to the
\texttt{git} version control system, which makes branch updates
trivial, and eases the process of bringing them back into the
mainline of the code.
With this in place, we are now able to implement our previous
aspirations: annual code releases, each accompanied by a light-weight
\textit{release} paper.

The present describes the 2023 release of \Cloudy{} (C23),
a major update to C17 in terms of atomic data, and it is
structured as follows.
Section~\ref{sec:access} describes the major changes that the project
has undergone these past few years.
Sections~\ref{sec:atomic} and \ref{sec:molecular} describe changes
to our atomic and molecular data.
Section~\ref{sec:grains} outlines recent improvements to our
treatment of grains.
Section~\ref{sec:xray} presents a summary of improvements to
\Cloudy's X-ray capabilities.
Section~\ref{sec:infra} describes changes to \Cloudy{} infrastructure,
including commands.
Finally, Section~\ref{sec:future} discusses current development
efforts that should be released in the next few years, as well
as our aspirations for future development.

\section{Online Migration}
\label{sec:access}

Fundamental changes to our infrastructure have occurred in the
last few years, most of them happening in Fall of 2019 and Fall
of 2020.
First, we were forced to move our user forum to a new website.
Then, in Fall 2020, our project was forced to vacate the servers that hosted
\url{https://www.nublado.org}, due to policy changes following the
acquisition of the host company by a third party.
The University of Kentucky has hosted our server since then.
The migration to a new server allowed us to
migrate to a more modern version control system, as well, namely
\texttt{git}.
As described below, one of us (JM) carried out the Fall 2020 migration.

\subsection{Migration of nublado.org}
There were few (if any) viable options for migrating the entire project to a new host without significant manual intervention.  That being the case, it was decided that this was an opportunity to migrate \Cloudy{} to more modern and flexible version control and tooling.
Trac\footnote{\url{https://trac.edgewall.org/}} is rather dated at this point.  It has not seen a significant update in many years, and it is built on Python 2.x, which is no longer being developed or supported.
Subversion\footnote{\url{https://subversion.apache.org/}} (SVN) is still actively maintained, but many development projects have moved to using Git, which has a larger community of users and developers.
After reviewing options for Git project hosting, it was ultimately decided to use GitLab\footnote{\url{https://about.gitlab.com/}} hosted at UK for the \Cloudy{} source code, issue tracker, and wiki.  GitLab offers a free, open-source edition and provides free access to their licensed features and support services for open-source research software.  

Migrating the existing data to GitLab in its entirety would be a difficult task.  The existing SVN repository contained several decades of revision history, and the Trac interface held a large number of wiki articles and issue data.  It was decided that the existing code revision history would not be migrated to Git.  Instead, there would be an initial commit in the Git repository containing a reference to the Subversion repository for historical purposes.  This greatly simplified the migration process.
Issue data and wiki pages were migrated to GitLab using TracBoat\footnote{\url{https://github.com/tracboat/tracboat}}.
This tool provided a very basic semi-automated migration but did not preserve all aspects of the content.
In particular, wiki links were removed and had to be re-added manually.  

The existing Trac site and Subversion repository were migrated to a new system at UK for historical reference purposes.  They were configured as read-only, and the default destination for project links is redirected to the GitLab instance.  Static files, including data sets, were also migrated and continue to be available and updated as needed.  

\subsection{User Forum Migration}

A forum where \Cloudy{} users can post questions or report problems
has been available since June 2005.
It had been hosted on \url{yahoo.com} until Fall 2019,
when the company decided to withdraw support for groups.
A key requirement for choosing a new host was to preserve the history of questions
and answers posted on \url{yahoo.com}.
\url{groups.io} met our needs, and the forum migrated to
\url{https://cloudyastrophysics.groups.io}.

The new platform has allowed for more versatility.
Our new setup now features a \texttt{Main} group, which preserves and extends our
Q\&A service.
It also carries an \texttt{Announcements} group, where important announcements,
e.g., about \Cloudy{} workshops, are made; a \texttt{Code} group, where users can
share scripts with other users; and finally a \texttt{Results} group, where users
can share results obtained using \Cloudy{} with the broader community.
Users are encouraged to subscribe to all these groups.


\section{Atomic Physics}
\label{sec:atomic}

\subsection{Upgrade to Chianti version 10}

\Cloudy{} has now adopted Chianti version 10.0.1.
Previously, the code had been using version 7.1, released in 2013.
The difficulty with upgrading earlier has been due to the changes the
database format since v7.1.
To remedy this, we developed a script to reprocess the version 10 data
into the version 7 format.
A detailed discussion of the script and changes to spectral line predictions
as a result of the new database is presented in \citet{CloudyChianti10}.
The reprocessing script has been made open-source and is available at
\url{https://gitlab.nublado.org/arrack}. 

Due to the large number of additional atomic levels in version 10, the full reprocessed Chianti database is $>$15 times the size of version 7. Since many of these levels are above the ionization limit of the corresponding species, we have omitted all auto-ionizing levels from the default version utilized by \Cloudy. Both the full reprocessed v10 database and the one without the auto-ionizing levels can be downloaded from \url{http://data.nublado.org/chianti/}. 

\subsection{Updates to the Stout database}

The format of the Stout \citep{Lykins2015} data files has been updated\footnote{For a full description
of the new format see \url{https://gitlab.nublado.org/cloudy/cloudy/-/wikis/StoutData}.}.
The most important change is that the spectroscopic information
must now be enclosed in double quotes in the \cdFilename{*.nrg} files.
This makes it easier for the code to extract this information which is now
included in the \cdCommand{save line labels} output. Also, the keywords in the
\cdFilename{*.coll} files have been updated to make parsing easier. The new
format is designated by the magic number ``17 09 05''.

The Stout database now supports having multiple atomic or molecular datasets
for a given species. This is necessary because it is not always possible to
unequivocally decide which calculation is the better one. In such a case, one
of these datasets will be designated as the default set, but the user has the
option to switch to a different set using the new option to the \cdCommand{species}
command.
For example, in the command \cdCommand{species ``Fe+6" dataset=``alt"},
\cdFilename{alt} is the nickname for the alternate dataset;
\Cloudy{} will use the data in the \cdFilename{Fe\_7\_alt.*} for this species.

\ion{Fe}{3} collision strengths are updated to \citet{BadnellBalance14}.
We had previously used data from \citet{Zhang1996}.
Energies are from NIST with missing levels taken from \citet{BadnellBalance14}.
\citet{2017ApJ...841....3L} describe how the data were matched to experimental energies.
Tests show that the total \ion{Fe}{3} cooling increases by nearly 50\%.

Certain ``baseline'' models
\citep[i.e., models without accurate collisions, see][]{2017RMxAA..53..385F}
in the Stout database have been
updated to use collisional, transition probability, and energy
data from the ADAS database.
NIST energies are employed for the lowest excited levels,
to permit the correct identification of spectral lines of
astronomical interest.
The species that have been updated include Mg$^{+9}$ \citep{2011A&A...528A..69L},
Al$^{+2}$ \citep{2009A&A...500.1263L}, and S$^{+4}$ \citep{2014A&A...572A.115F}.

\subsection{Fe II}

The $\feii$ ion has a complex
structure with 25 electrons, and 
is a ``grand challenge'' 
problem in atomic physics.
An accurate set of radiative 
and collisional atomic data is 
therefore needed 
to treat selective excitation, 
continuum pumping, and 
fluorescence, which are known to 
be crucial for the $\feii$ emission 
\citep[e.g.,][]{2000ApJ...543..831V,2004ApJ...615..610B,Bruhweiler_2008,2012MNRAS.425..907J,2016ApJ...824..106W,Netzer2020}.
Uncertainties in the atomic data 
have been a longstanding limitation 
in interpreting line intensities.

\begin{figure}[t]
    \centering
    \includegraphics[width=\columnwidth]{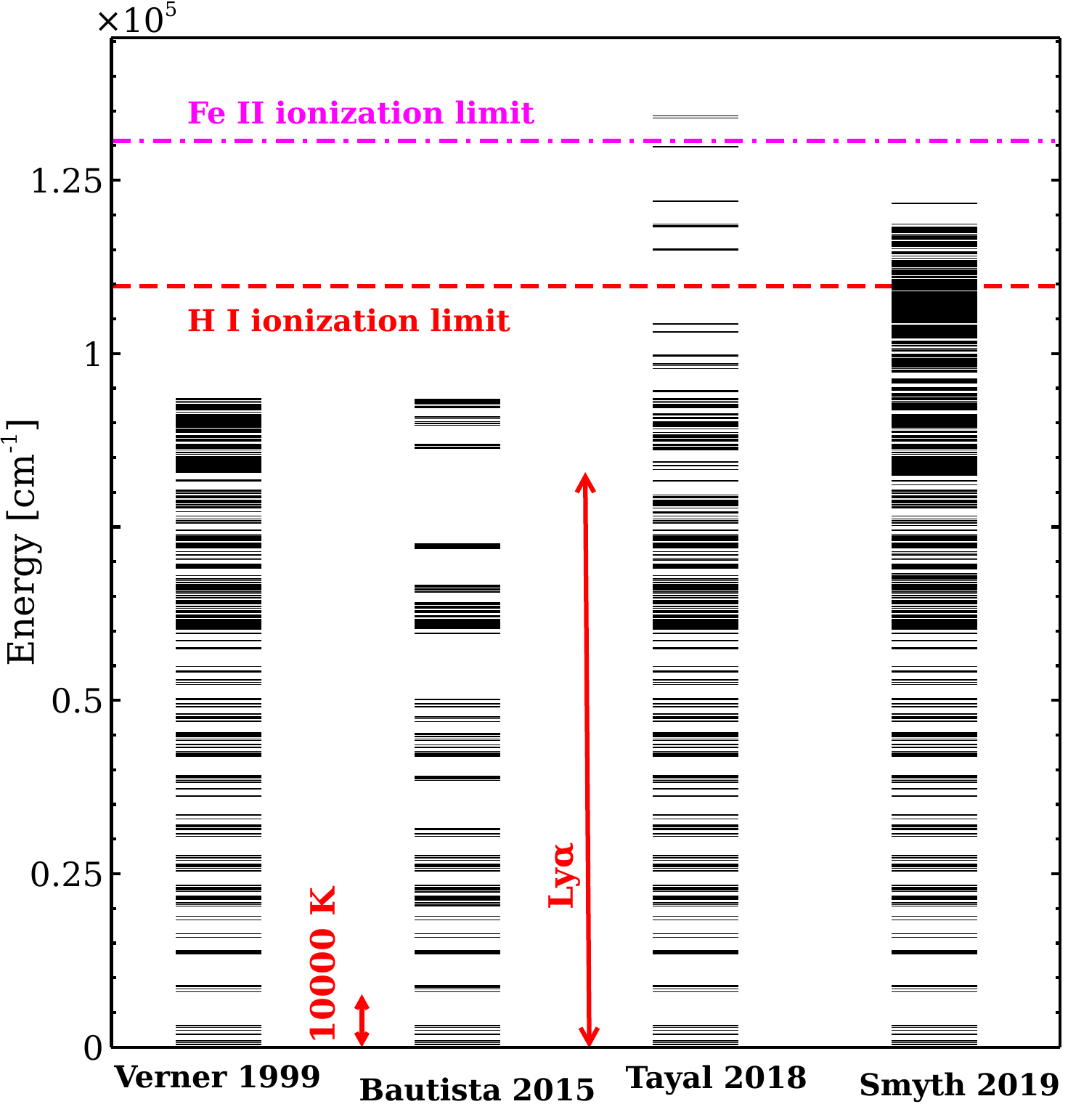}
    \caption{Energy level structure of the $\feii$ models available with \Cloudy{}.
    From left to right, the atomic datasets are those of \citet{1999ApJS..120..101V}, 
    \citet{2015ApJ...808..174B}, \citet{2018PhRvA..98a2706T}, and
    \citet{2019MNRAS.483..654S}. 
    The $\hi$ and $\feii$ ionization limits, the Ly$\alpha$ energy, an important
    source of photoexcitation for $\feii$, and the thermal energy corresponding to
    10$^4$~K are also indicated.
    Adapted from \citet{CloudyFeII2021}.}
    \label{fig:feii_atomic}
\end{figure}

Until recently, \Cloudy{} shipped with the \citet{1999ApJS..120..101V} model,
which has 371 level that reach 11.6~eV, has about 68,000 transition probabilities,
but uses the ``g-bar'' approximation for collision strengths.
Due to its limitations, \citet{CloudyFeII2021} explored three other atomic models,
that are now available in the $\Cloudy$; their energy levels are compared in  
Figure \ref{fig:feii_atomic}.

Of relevance to the 2000--3000\AA\ ultraviolet range are the datasets of
\citet{2018PhRvA..98a2706T} and \citet{2019MNRAS.483..654S}.
The \citet{2018PhRvA..98a2706T} model has 340 energy levels with the highest 
energy of $\sim$ 16.6 eV, that is, it goes above the ionization limit (16.2 eV).
However, its density of states in high-lying energy levels is low, 
as shown in Figure \ref{fig:feii_atomic}.
Transitions between these energy levels produce about 58,000 emission lines 
with uncertainties in transition probabilities of $\lesssim$ 30\%
(in the 2200\AA--7800\AA). 
On the other hand, the \citet{2019MNRAS.483..654S} model includes 716 
levels in the close coupling (scattering model) calculation, with the
highest energy level reaching 26.4 eV.
These levels produce about 256,000 emission lines. 
The \citet{2019MNRAS.483..654S} dataset also contains autoionizing levels, 
but its density of states in the high-lying energy states is larger than the 
\citet{2018PhRvA..98a2706T} dataset.
Further details on these atomic models can be found in \citet{CloudyFeII2021}.
Both models have been incorporated into \Cloudy{}.

\begin{figure}
\centering
\includegraphics[width=1\columnwidth]{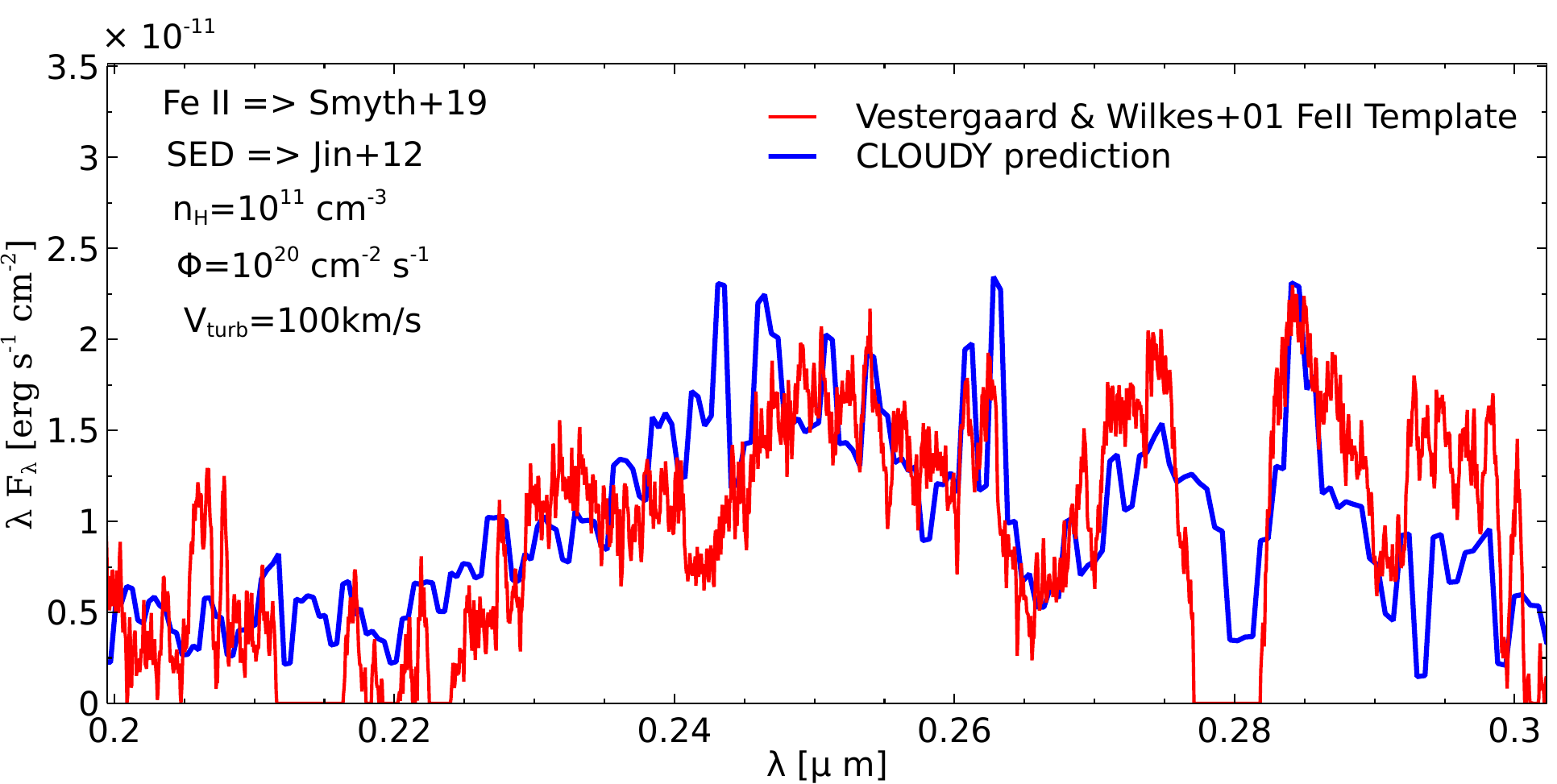} 
\caption{Comparison between the observed $\feii$ UV template of \citep{2001ApJS..134....1V} 
         and the $\Cloudy$ predicted $\feii$ UV spectrum with V$_{\rm turb}$ = 100 km/s.}
\label{fig:w_wo_turb}
\end{figure}

\citet{CloudyFeII2021} showed that with the 
\citet{2019MNRAS.483..654S} dataset,
\Cloudy{} produces a spectrum that is in satisfactory agreement 
with the template spectrum of the I~Zw~1 Seyfert galaxy
\citep{2001ApJS..134....1V}.
Briefly, the better agreement of the Smyth dataset is due to the higher density of
highly excited states, which enhance the effects of continuum fluorescence and lead
to brighter emission lines at short wavelengths.
Further details are described in the paper.
Figure~\ref{fig:w_wo_turb} illustrates the quality of that comparison.

\subsection{K$\alpha$ energies}

\begin{figure}
\centering
\includegraphics[width=\columnwidth]{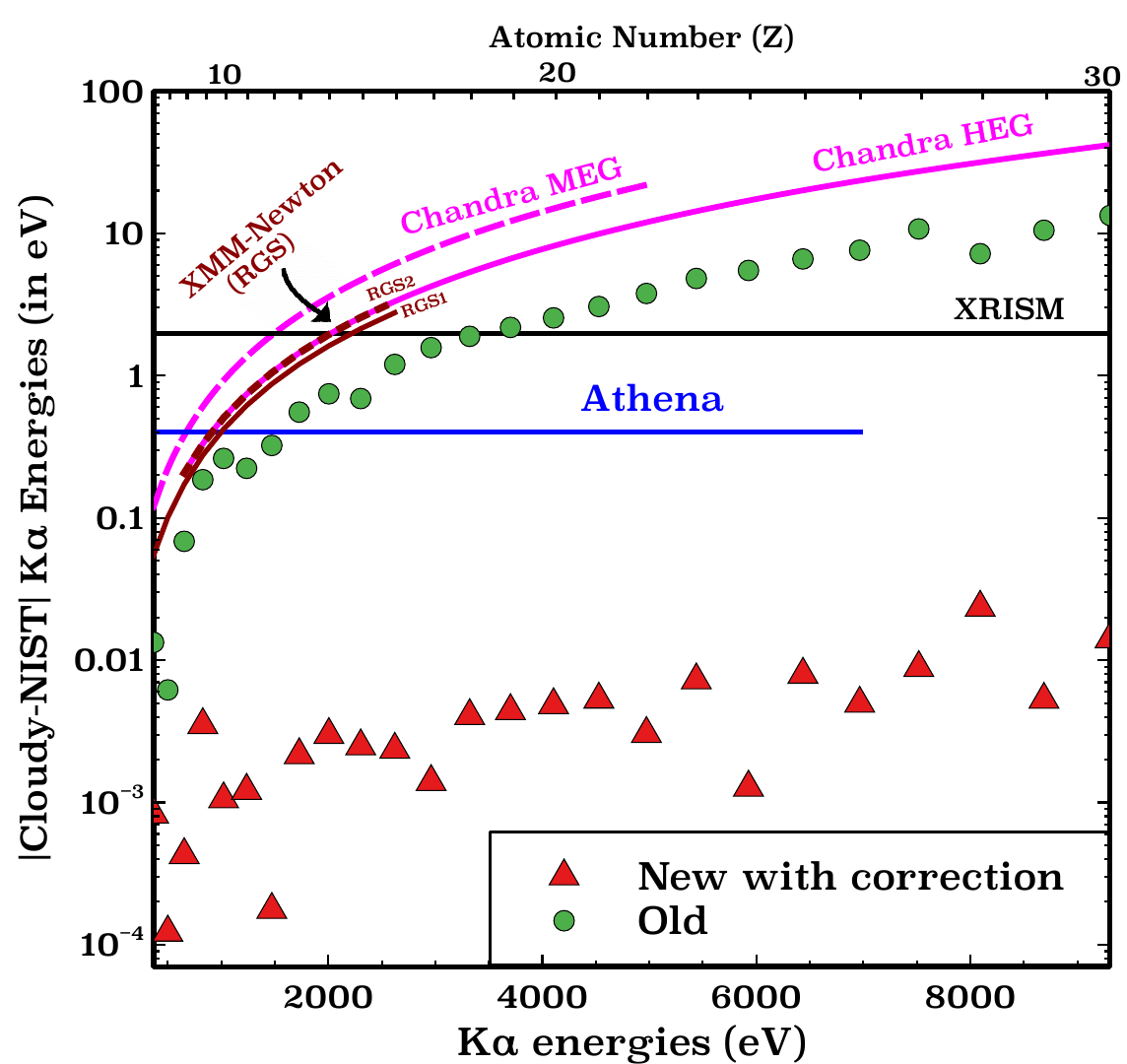}
\caption{The absolute value of the difference between NIST and {\Cloudy} K$\alpha$ energies
versus K$\alpha$ energies for H-like ions of elements between $6 \leq Z \leq 30$.
Red triangles show the difference between energies in the current version (as of C17.03)
of {\Cloudy} and NIST, while green circles show the same difference for previous
versions of {\Cloudy} (from $\sim$2005 to C17.02).}
\label{f:nist}
\end{figure}

We incorporated precise  H-like K$\alpha$ energies for elements between 6 $\leq$ Z $\leq$ 30
to match laboratory energies \citep{KaCorr}. Figure \ref{f:nist} compares the difference in  K$\alpha$ energies between NIST \citep{2018APS..DMPM01004K}
and {\Cloudy} for the updated {\Cloudy} energies and the old {\Cloudy} energies appearing 
in C17.02. The revised K$\alpha$ energies are $\sim$ 15-4000 times more precise than that of C17.02. This energy precision is also much superior to the energy accuracy of the current and future X-ray instruments like \textit{Chandra}, \textit{XMM-Newton}, \textit{XRISM}, and \textit{Athena}, as shown in Figure  \ref{f:nist}.
The improved {\Cloudy} energies will therefore be in  excellent agreement with the future microcalorimeter observations.\\

\subsection{Inner shell energies}

We updated the fluorescence K$\alpha$ energies of Si \textsc{ii-xi} and S \textsc{ii-xiii} used in \Cloudy{} with the experimental data reported in \citet{2016ApJ...830...26H}.
The energies of the lines in the past versions of \Cloudy{} are mainly taken from Table 3 of \citet{1993A&AS...97..443K}, which contains the fluorescence yields, energies and Auger electron numbers for elements and ions from Be to Zn. Even though these values were in accordance with the theoretical calculations available at the time of the publication, today, this data-set is not accurate enough to model some already available high-resolution spectra \citep{2021A&A...648A.105A,2021RNAAS...5..149C}, and certainly for future X-ray spectra having eV resolution.

\begin{figure}[!t]
  \includegraphics[width=\columnwidth]{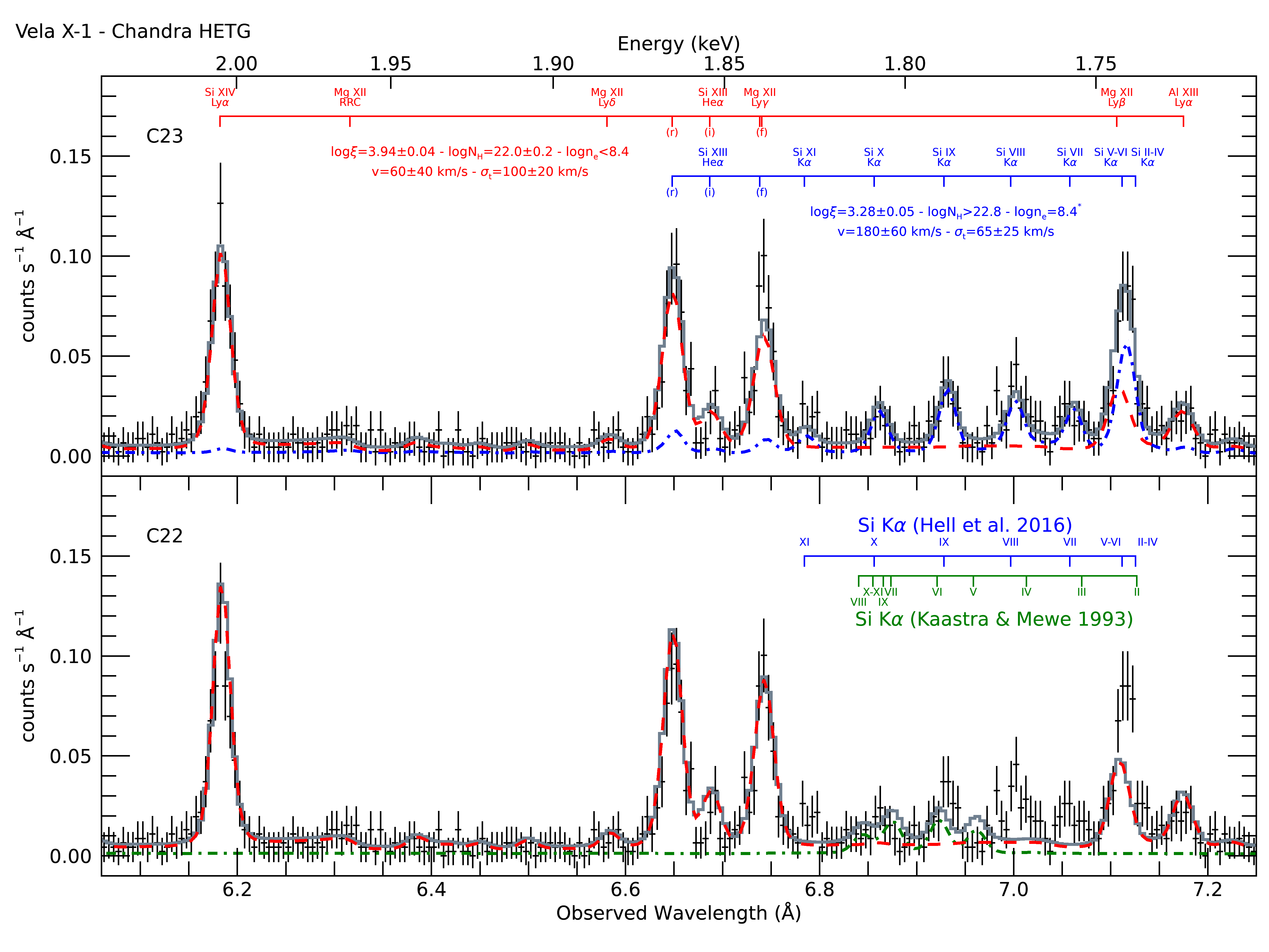}
  \caption{Visually co-added \textit{Chandra} MEG $\pm1$ order spectra of the HMXB Vela X-1 at the orbital phase $\phi_{orb}=0.75$ \citep[see][]{2021A&A...648A.105A}. \textit{Top}: Best fit model with \Cloudy{} (grey solid line), using the improved energies for the Si fluorescence lines, available in C23. The specific contributions of each gas component are labeled (red dashed line and blue dot-dashed line), together with the best-fit parameters and 90\% confidence level uncertainties \citep[see][for details]{2021RNAAS...5..149C}. \textit{Bottom}: As above, but with the previous version of \Cloudy{}, C17. The low ionization component is here labeled in green, together with the adopted Si K$\alpha$ lines \citep[from][]{1993A&AS...97..443K}. For ease of comparison, the improved energies from \citet{2016ApJ...830...26H} are in blue, as in the top panel. Adapted from \citet{2021RNAAS...5..149C}.}
  \label{fig:vela}
\end{figure}

For this reason, we updated the energies from O-like to Be-like ions of Si and S (Si \textsc{vii-xi} and S \textsc{ix-xiii}) adopting the centroids for unresolved blends given in Table 3 of \citet{2016ApJ...830...26H}. Regarding the low-ionization lines, individual energies are taken from their Table 5, where the values of Si \textsc{ii-iv}, Si \textsc{v-vi}, S \textsc{ii-vi} and S \textsc{vii-viii} are listed. K$\alpha_{1}$ and K$\alpha_{2}$ are not experimentally resolved in \citet{2016ApJ...830...26H}, and their difference in energy is lower than the expected resolution of future X-ray microcalorimeters, so we assumed the same energy for both.  \citet{2021RNAAS...5..149C}  provided a demonstration of the impact that such an update can have on the high-resolution spectra of the High Mass X-ray Binary Vela X-1 (see Figure \ref{fig:vela}). 


\subsection{$\ell$-changing collisions}

Momentum change collisions by protons were deeply revised in C17. An upgrade of
the theory of \citet{PengellySeaton1964} (hereafter PS64), dubbed PS-M, was
published by
\citet{Guzman.II.2017}, correcting the results at high density and low temperature of PS64
and getting a better agreement with the quantum-mechanical \textit{ab initio} results
from \citet{Vrinceanu2001} (hereafter VF01).
The theory has been further refined by \citet{Badnell-lchanging}, now called PS-M20.
PS-M/PS-M20 results were in better
agreement with the quantal calculations than the semi-classical calculations
from \citet{VOS2012} (hereafter VOS12), which underestimate VF01 by a factor
$\sim6$ \citep{Guzman.I.2016}. Semi-classical rates were corrected by
\citet{Vrinceanu2017} to get
agreement with VF01 and PS-M. \citet{Vrinceanu2019} report in their figure 2 a
disagreement of PS-M rates with quantum mechanical rates for high $n$. However,
we have confirmed that PS-M rates actually agree with the quantum mechanical
ones for the results plotted in their figure. Their reported
disagreement can be explained because they calculated the excitation
$\ell\to \ell+1$ collisions incorrectly using the formula given by
\citet{Guzman.II.2017}, optimized for de-excitation collisions ($\ell\to \ell -1$). If
micro-reversibility is applied to the results of \citet{Vrinceanu2019} the
obtained rates agree with the quantum ones for all the range of the figure
\citep{Badnell-lchanging}. 

\citet{Guzman.2017.TwoPhoton} compared in their tables 1 and 2 PS-M effective recombination
rates to $n=2$ levels of hydrogen with the ones quoted in tables 4.10 and 4.11 of \citet{AGN3},
obtained from \citet{PengellySeaton1964}.
PS-M produces effective recombination coefficients to $2s$ that are bigger in a $\sim0.6$\%
for case A and a $\sim0.1$\% for case B, while recombination to $2p$ is smaller by $\sim1-2$\%.
Similarly, they found an agreement up to $\sim5$\% for the emissivities of
$2s\,^2\text{S} \to 2p\,^2\text{P}_{3/2}^o $ and up to a $\sim0.5$\% for
$2s\,^2\text{S} \to 2p\,^2\text{P}_{3/2}^0 \to 2p\,^2\text{P}_{1/2}^o $.
We do not expect this to influence Ly$\alpha$ or the two-photon emission spectrum.

PS-M20 theory has been implemented in the latest versions of \Cloudy{} for H-like and He-like ions.
\Cloudy{} selects PS-M20 theory by default if not specified.
The command that selects the new PS-M20 results is: 
\begin{verbatim}
database H-like collisions l-mixing 
                            Pengelly PSM20
\end{verbatim}
for H-like collisions, and 
\begin{verbatim}
database He-like collisions l-mixing PSM20   
\end{verbatim}
for He-like systems.

\subsubsection{The importance of a correct number of resolved levels: printing $\ell$-changing critical densities}

Accurate modeling of recombination lines for H-like and He-like ions
requires a good description of the ion energy levels.
\Cloudy{} distinguishes between angular momentum $\ell$-{\it resolved} levels
and {\it collapsed} levels, for which $\ell$-levels are populated according to
their statistical weight \citep[see][for more details]{2013RMxAA..49..137F}.

Critical densities are defined as the density where collisional rates equal
radiative de-excitation, $n_{e,c} \, q_{lu}=\tau^{-1}_{ul}$
\citep{PengellySeaton1964}, where $\tau_{ul}$ is the half-life of radiative
de-excitation between $u$ and $\ell$ sub-shells, and $q_{lu}$ is the effective
coefficient of collisional excitation between $\ell$ and $u$ sub-shells.
Then averaging over $\ell$ we can obtain the critical density for the shell $n$:
\begin{equation}
    n_{e,c} = \frac{1}{q_{n}\tau_{n}},
    \label{eq:critdens}
\end{equation}
For densities above $n_{e,c}$, collisions will be faster than radiative decay and dominate, ultimately bringing the $n$-shell to be statistically populated in its $\ell$-sub-shells.

\Cloudy{} now has a new command,
\begin{verbatim}
print critical densities
\end{verbatim}
that can be used to print $\ell$-changing critical densities for H-like
and He-like ions in the output file.
This command aims to help choose a physically-motivated number
of resolved levels to employ in a simulation for each ion.

Optionally, \verb+H-like+ or \verb+He-like+, can be added together with an element name to specifically print the critical densities for an ion. For example, while the command line above prints critical densities for all resolved levels included in the simulations for all H-like and He-like ions, the more specific command,
\begin{verbatim}
print critical densities H-like
\end{verbatim}
prints only critical densities for H-like ions, while
\begin{verbatim}
print critical densities H-like helium
\end{verbatim}
prints only critical densities for the He$^+$ ion.

\begin{figure}
  \includegraphics[width=\columnwidth]{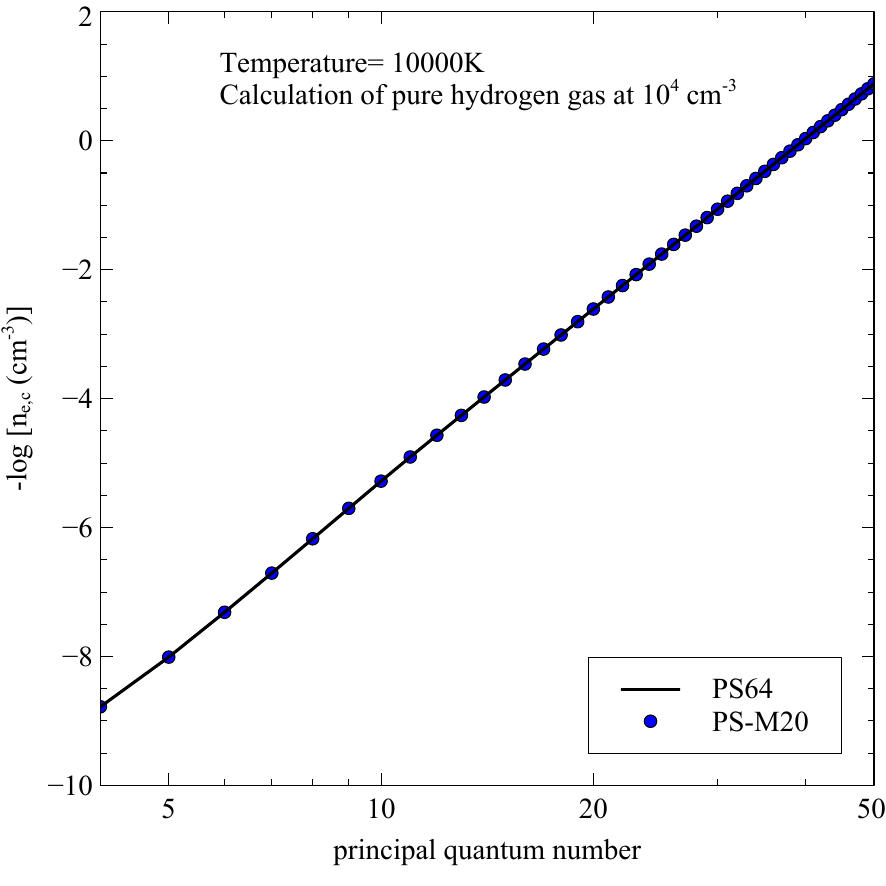}
  \caption{\label{fig:PS64} Comparison of critical densities from equation
           \ref{eq:critdens} with $\ell$-changing collisions from PS-M20 with
           the results of figure 4 of \citet{PengellySeaton1964}.
           We have chosen a pure hydrogen gas at electron temperature $T=10000$K
           and electron density $n_e=10^4\pcc$. The agreement is complete.}
  \end{figure}

In Fig. \ref{fig:PS64} we have plotted the critical densities obtained with
\Cloudy{} using PS-M $\ell$-changing theory versus the principal quantum number.
Critical densities from figure 4 of \citet{PengellySeaton1964} are also plotted for comparison.
Complete agreement is shown in the Figure.
When densities are higher than the critical densities, collisions dominate and
the $\ell$ quantum numbers redistribution is faster.
At much higher densities the population of the $\ell$-sub-shells will be statistical.
Figure \ref{fig:PS64} can be used to determine the number of levels that should be
treated as resolved in $\ell$.
As a rule of thumb, it is safe to add ten units to the principal quantum number for
which the density is critical to ensure that collisions will statistically populate
the levels treated as collapsed.
For example, in a simulation of a H~II region with density $n_e=10^4\pcc$,
the principal quantum number corresponding to that density would be $n\sim 15$,
according to Fig \ref{fig:PS64} (we can use the \verb+print critical density+ command
to verify the critical densities for all resolved levels).
A safe number of resolved levels to use is all levels $n\leq25$, that can be included
in the simulations with the line \citep{2017RMxAA..53..385F}:
\begin{verbatim}
database H-like hydrogen levels resolved 25
\end{verbatim}

Note that different conditions of electron temperature and densities might cause the
critical densities vary, obeying the temperature dependence of the $\ell$-changing cross
sections, as well as their dependence on density due to the Debye cut-off of the collision
probabilities \citep{PengellySeaton1964,Guzman.I.2016}.
 
\subsection{n-changing collisions}

Principal quantum number-changing electron collision data 
were analyzed by \citet{Guzman.III.2019}.
C17 and previous versions used semi-empirical data from 
\citet{Vriens1980} (hereafter VS80).
\citet{Guzman.III.2019}  suggested using the semiclassical
straight trajectory Born approximation of
\citet[hereafter LB98]{Lebedevandbeigman1998}.
The latter is within a factor of $\sim2$ of 
VS80 collisions and have the same dependency on the high- and 
low-energy range.
Care must be taken when dealing with highly charged ions as the
straight trajectory approximation would fail, especially at low energies,
producing an underestimation of the rates.
In that case, it would not be safer to use VS80, as this is intended only for atoms.
Further theoretical work is needed for a better theory for highly charged ions.

While LB98 is the default theory for both H-like and He-like 
$n$-changing collisions, it is possible to choose between 
different theories in \Cloudy{} using the command:
\begin{verbatim}
database H-like collisions Lebedev
\end{verbatim}
where \verb+H-like+ can be modified to \verb+He-like+ and the options are:
\begin{itemize}
\item \verb+Lebedev+ (default) for semiclassical straight trajectory theory \citep{Lebedevandbeigman1998}.

\item \verb+Vriens+ for \citet{Vriens1980} semi-empirical approximation.

\item \verb+Fujimoto+ for the semi-empirical formula fit of \citet{Fujimoto1978}.

\item \verb+van regemorter+ for the averaged gaunt factor formula proposed by \citet{VanRegemorter1962}.
\end{itemize}
A comparison and analysis of these theories and application to different cases can be found at \citet{Guzman.III.2019}.

\subsubsection{Masing of H$n\alpha$ lines}

In contrast to C17, masing of hydrogen lines is now allowed.
\citet{Guzman.III.2019} predict masing of H$n\alpha$ radio recombination lines
for low-density clouds ($n_e \leq 10^8 \pcc$).
These authors also predict masing of the H$n\alpha$ lines, with $n$ ranging
between 50 and 190, for a model of the Orion Blister.
However, the number of masing lines decreases for data sets other than LB98.
In these cases, the higher collisional rates bring the populations of the Rydberg
levels closer to LTE, thus suppressing masing.

\section{Molecular Data}
\label{sec:molecular}

\subsection{\htwo{}}

A large model of molecular hydrogen was introduced by \citet{Shaw2005}.
The level energies, which we use to derive line wavelengths, have been updated
to \citet{Komasa16H2Eneries}. 
This work incorporates many high-order effects in the \htwo\ wavefunctions,
which result in $\sim 1$ wavenumber changes in level energies.
We derive line wavelengths from these energies, so small changes in wavelength result.
The \citet{Komasa16H2Eneries} energies are thought to be a significant improvement over previous data sets.

The previous version of \Cloudy\ included the \citet{Lique15} H -- \htwo\ collisional data as an option,
although they were not used by default.  We now use this as our preferred H -- \htwo\ collision data set.
Compared with previous calculations, these data extend to higher vibrational manifolds and include
ortho-para changing reactive collisions.
Tests show that the \htwo\ 2.121 \micron\ line intensity changes by roughly 50\%, becoming stronger
in some PDR sims.

\subsection{Other molecules}

Molecular lines are sensitive to underlying physical conditions. 
Hence, they reveal physical conditions in various astrophysical environments 
when interpreted correctly. It is always our aim to predict more molecular 
lines with better precision. We do it in two ways. Firstly, by including 
more molecules in the \Cloudy{} molecular network. Secondly, by updating the 
existing molecular network. Below, we mention such recent efforts.

\begin{figure}[!t]
  \includegraphics[width=\columnwidth]{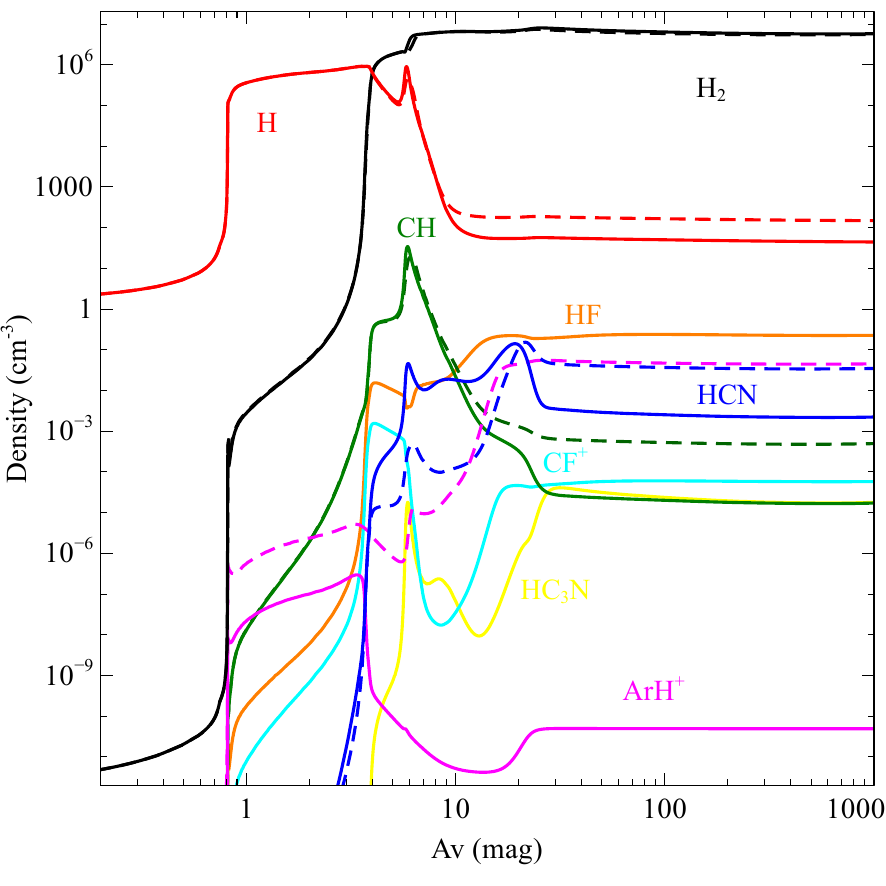}
\caption{Variation of densities of a few molecules as a function of $A_{\rm V}$ for a H~II and PDR model ("h2$_{-}$orion$_{-}$hii$_{-}$pdr.in'' from the \Cloudy{} download under the directory \texttt{tsuite}). 
The name of each molecule and the line representing its density are depicted in the same color. 
The solid lines represent simulations using this version, and the dashed lines represent simulations 
using an earlier version C17.}
\label{fig:structure_plot}
\end{figure}

\citet{2022ApJ...934...53S} have included the gas-phase energy levels, 
radiative and collisional rates for HF, CF$^+$, HC$_3$N, ArH$^+$, HCl, HCN, CN, CH, 
and CH$_2$ into \Cloudy{}'s molecular network.
The energy levels and collisional rate coefficients were taken from the
upgraded LAMDA Database \citep{2020Atoms...8...15V}.
However, reaction rates were from the UMIST Database for Astrochemistry
(UDfA 2012; RATE12), specifically, \citet{2014A&A...566A..30R,2014A&A...566A..29S,2017MNRAS.472.4444P}. 
As a result, \Cloudy{} now predicts the line intensities and column densities
of these molecules in addition to those included in the previous version. 
Figure \ref{fig:structure_plot} \citep{2022ApJ...934...53S} shows the variation of densities of a few molecules 
as a function of $A_{\rm V}$. The name of each molecule and the line representing its density are depicted in the 
same color. The solid lines represent simulations using this version, and the dashed lines represent simulations 
using the earlier version C17. 

\begin{figure}[!t]
  \includegraphics[width=\columnwidth]{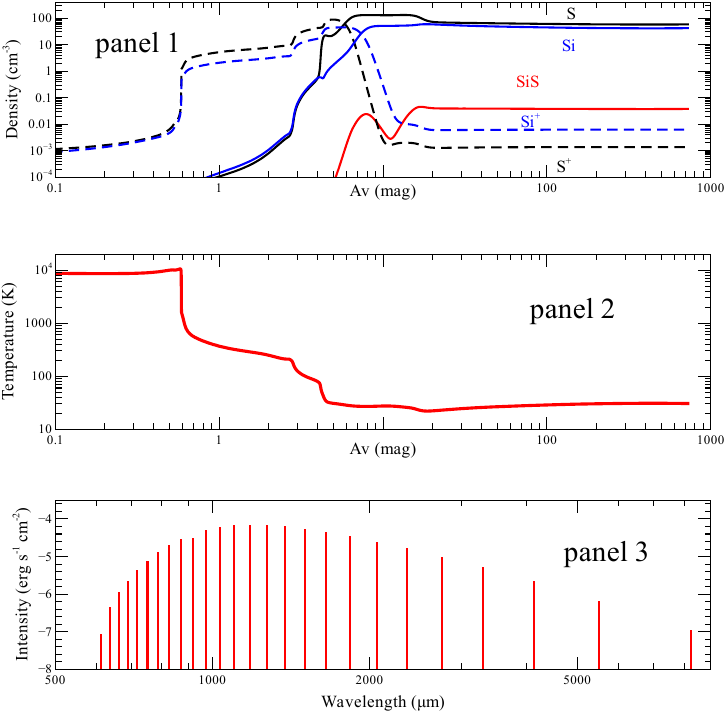}
\caption{Panel 1: Variation of SiS density as a function of $A_{\rm V}$ for an H~II and
         PDR model (``h2$_{-}$orion$_{-}$hii$_{-}$pdr.in'' from the \Cloudy{} download
         under the directory \texttt{tsuite}).
         Panel 2: Variation of temperature across the cloud.
         Panel 3: Model predicted intensities of various SiS rotational lines.}
\label{fig:SiS}
\end{figure}

Likewise, we have included the gas-phase chemical reactions, energy levels, and
radiative and collisional rates of the SiS molecule \citep{2023RNAAS...7...45S}. 
The energy levels, Einstein's radiation coefficients and collisional rate coefficients
with H$_2$ molecules were taken from the upgraded LAMDA database.
In addition, we have included collisions with H \citep{ANUSURI2019e01647}
and He \citep{{2007A&A...472.1037V},{2008JPhB...41o5702T}}.
The chemical reaction rates were from various sources, UDfA, namely
\citet{2018ApJ...862...38Z}, \citet{1998A&A...330..676W}, \citet {2021SciA....7.7003D};
and the Kinetic Database for Astrochemistry,\footnote{\url{https://kida.astrochem-tools.org/}} respectively.
Figure \ref{fig:SiS} \citep{2023RNAAS...7...45S} demonstrates predicted intensities of various rotational lines of SiS for a H~II and PDR model (``h2$_{-}$orion$_{-}$hii$_{-}$pdr.in'') from the \Cloudy{} download under the directory \texttt{tsuite}.

\citet{2023RNAAS...7...45S} included only rotational levels of SiS. However, we have now included the vib-rotational levels (private communication 
with Ziwei Zhang). 

Any species' predicted column densities and line intensities depend on rate coefficients. 
We use mostly UDfA rate coefficients.  In the UDfA, a two-body gas-phase chemical reaction rate coefficient 
$k$ ($\rm{cm^3 \, s^{-1}}$) is given by the usual Arrhenius-type formula,
\begin {equation}
k=\alpha \left(\frac{T}{300}\right)^\beta \, \exp(-\gamma /T),
\end {equation}
where $T$ is the gas temperature. 
Reactions with $\gamma$ $<$ 0 will become unphysically large at low temperatures. \citet{2011A&A...530A...9R} 
has addressed the divergence of rate coefficients at low temperatures. 
A similar problem occurs for high temperatures encountered with CLOUDY.
We apply a temperature cap T$_{cap}$ for $\beta$ $>$0 to avoid this. 
For $T > T_{cap}$, the rate coefficients retain the same values as 
at T$_{cap}$. Though ad hoc, we choose T$_{cap}$ = 2500K \citep{2023RNAAS...7..153S}.
This affects the warm, 5000K - 10000K, collisionally ionized clouds.

Cosmic-ray ionization rate plays an important role in ISM and is an active field of research.
\citet{2021ApJ...908..138S} demonstrated that the abundance of PAHs affects the free electron density,
which changes the  H$_3^+$ density and hence the derived cosmic-ray ionization rate of hydrogen.
We suggested that for the average Galactic PAH abundance, the cosmic-ray ionization rate of
atomic hydrogen to be 3.9$\pm$ 1.9\e{-16} \ps.
The command
\begin{verbatim}
cosmic rays background 1.95 linear
\end{verbatim}
sets this rate.
Furthermore, we showed that the cosmic-ray ionization rate of hydrogen derived using H$_3^+$
is much higher when PAHs are absent.

Furthermore, \citet{2020RNAAS...4...78S} updated the mean kinetic energy of the secondary electrons (from 20eV to 36eV) produced by cosmic rays. This affects the cosmic-ray dissociation of molecular hydrogen and dense cloud chemistry.

\section{Grain Data}
\label{sec:grains}

\subsection{New Refractive Index Files}

New refractive index files have been added for astronomical silicate and
graphite using the data described in \citet{Draine2003}. 
These add much
more structure to the inner-shell photoionization edges seen
in the X-ray regime.

\subsection{Interstellar Grain Absorption}



We have introduced a new option to the \verb+metals deplete+ command
to use the more self-consistent depletion pattern described in \citet{Jenkins2009}.
An in-depth discussion on this depletion pattern and how it affects the predicted
spectra is presented in \citet{CloudyGrainDepl, Gunasekera2023}.
This new element-selective depletion option can be enabled with the
\begin{verbatim}
metals deplete Jenkins 2009
\end{verbatim}
command.
The depletion parameters $A_X$, $B_X$, and $z_X$, specific for each element X,
are read in from an external file called \cdFilename{Jenkins09\_ISM\_Tab4.dep}, and are
used to compute the depletion scale factor $D_x$ using the equation
\begin{equation}
	D_x = 10^{B_X+A_X(F_*-z_X)},
	\label{eq:Jenkins_Dx}
\end{equation}
where $F_*$ represents the degree of depletion across all elements.
This $D_x$ factor then multiplies the reference abundance to produce the
post-depletion gas-phase abundances.
By default, $F_*=0.5$.
However, its value may be adjusted with the command
\begin{verbatim}
metals depletion jenkins 2009 fstar <value>
\end{verbatim}
where the \verb+<value>+ must range between 0 and 1.

An analysis of strong spectral-lines 
(log([O\,{\sc iii}] $\lambda$5007/H$\beta$), log([N\,{\sc ii}] $\lambda$6583/H$\alpha$), log([S\,{\sc ii}] $\lambda\lambda$6716,6731/H$\alpha$), and log([O\,{\sc i}] $\lambda$6300/H$\alpha$)) 
from a \Cloudy{}  model of a generalized H II region,
based on SDSS-IV MaNGA observations \citep{bundy2015,yan2016b},
revealed that varying $F_*$ affects the spectral line intensities
and the thermal balance of the ionized cloud \citet{Gunasekera2023}.
The user must alter the grain abundance to match the degree of depletion ($F_*$).
To do this, compute the fraction of the total abundance of heavy elements locked
in dust grains in the given $F_*$ relative to the total abundance of heavy elements
locked in dust grains at $F_*=0.5$,
\begin{equation}
	grains = \frac{\sum_X{(X_{dust}/H)_{F_*}}}{\sum_X{(X_{dust}/H)_{0.5}}}.
\end{equation}  
This fraction can then be given in the \texttt{grains} command to change the dust abundance self-consistently.

\section{X-ray Predictions}
\label{sec:xray}

\subsection{Microcalorimeters}

Historically, \Cloudy{} made X-ray predictions but was not designed
for high-resolution spectral analysis.
In preparation for the upcoming microcalorimeter missions \textit{XRISM}
and \textit{Athena}, we have extended \Cloudy{} to make it compatible
with high-resolution spectral analysis in the X-ray regime.
\citet{Chakraborty.I} demonstrated the effects of Li-like iron on the
Fe XXV K$\alpha$ line intensities via Resonant Auger Destruction
\citep[RAD; e.g.,][]{1978ApJ...219..292R,  1996MNRAS.280..823M, 2005AIPC..774...99L}.
Although initially motivated by the Perseus cluster, this analysis was extended
to include a wide range of column densities encountered in astronomy.

We also showed line-broadening effects produced by electron scattering.
The command \verb+no scattering escape+ was introduced to ignore scattering
of photons off of thermal electrons \citep{Chakraborty.I}.
When line photons scatter off high-speed electrons, a fraction of them
receive large Doppler shifts from their line center, creating super-broad
Gaussian profiles. 
Such broad line profiles will not be detected in future high-resolution X-ray telescopes.
The purpose of the above command is to model the unscattered photons
that be observed by these future X-ray missions.
The command \verb+no absorption escape+ was introduced to ignore
absorption by background opacities.

Comparing the observed spectra by \textit{Hitomi} with \Cloudy{} simulated spectra,
\citet{Chakraborty.II} presented a novel diagnostic for measuring column densities
transitioning from optically thin (Case A) to optically thick (Case B) in H- and He-like iron.
The effects of metallicity  and turbulence on Fe XXV K$\alpha$ line ratios were also
demonstrated using the Perseus cluster as a reference.

We have also updated the collision strengths of the Fe K$\alpha$ lines  using recent
calculations by \citet{2017A&A...600A..85S}, replacing the old collision rates by
\citet{1987ApJS...63..487Z}.
The new rates are calculated based on the independent process and isolated resonance
approximation using distorted waves (IPIRDW) technique.
Updates to the collision rates resulted in significant differences in the estimated
Fe K$\alpha$ line ratios, as described in \citet{Chakraborty.II}.

\citet{Chakraborty.III} extended the classic Case A and Case B \citep{2006agna.book.....O}
to less familiar regimes Case C and Case D in the X-ray band.
Previous works on these limits focused on the optical, ultraviolet, and infrared regimes
\citep{1937ApJ....85..330M, 1938ApJ....88...52B, 1953ApJ...117..399C,  1999PASP..111.1524F, 2017PASP..129h2001P},
but X-ray wavelengths were rarely studied.
The net X-ray spectrum for all four cases within the energy range 0.1-10 keV was simulated
at the resolving power of \textit{XRISM}.

\citet{Chakraborty.IV} demonstrated atomic processes modifying soft X-ray spectra.
This includes the enhancement in line intensities via continuum pumping in photoionized
environments, and suppression in line intensities through photoelectric absorption and
electron scattering in collisionally-ionized and photoionized enviromnents.
A hybrid of \Cloudy{} simulated collisionally-ionized and photoionized model was used to
fit the \textit{Chandra} Medium Energy Grating (MEG) spectrum from V1223 Sgr,
an intermediate polar.
This was the first application of the new \Cloudy{} interface compatible with
high-resolution spectral analysis in the X-ray regime.

We also increased the default number of levels in our default instance of the Fe$^{16+}$ ion. 
The default limit to the
number of its levels  suppressed the 15.013~\AA{} line, which is prominent in
soft X-ray spectra.
\citet{2017RMxAA..53..385F} has an extensive discussion of our choice of a default
number of levels, the effects on a calculation, and how to change it.

Finally, we have improved the energy grid resolution of \Cloudy{}'s coarse continuum
to better suit it for tailoring models to the upcoming \textit{XRISM} mission.

\subsection{Inner Shell Ionization}

The X-ray portion of most SEDs has little effect on the ionization 
of an ionized Cloud, as discussed for AGN in the first appendix
of \citet{2023MNRAS.523..646T}.
A more general discussion is presented here.

The photoionization rate for a given shell $n$ is
\begin{equation}
\Gamma_n = \int_{\nu_0}^{\infty} \sigma_{\nu} \ \phi_{\nu} \ d\nu\ \  [{\rm s}^{-1}]
\label{eq:Gamma}
\end{equation}
where $\nu_0,\ \sigma_{\nu}$ are the photoionization threshold of shell $n$ [Ryd] 
and the cross section [cm$^{-2}$], respectively, and $\phi_{\nu}$ is the flux of
ionizing photons [cm$^{-2}$ s$^{-1}$ Ryd$^{-1}$].  
The total photoionization rate is the sum over all shells is
\begin{equation}
\Gamma_{total} = \Sigma_n \ \Gamma_n \  \ \ [{\rm s}^{-1}]
\end{equation}

\begin{figure}
\centering
\includegraphics[width=\columnwidth]{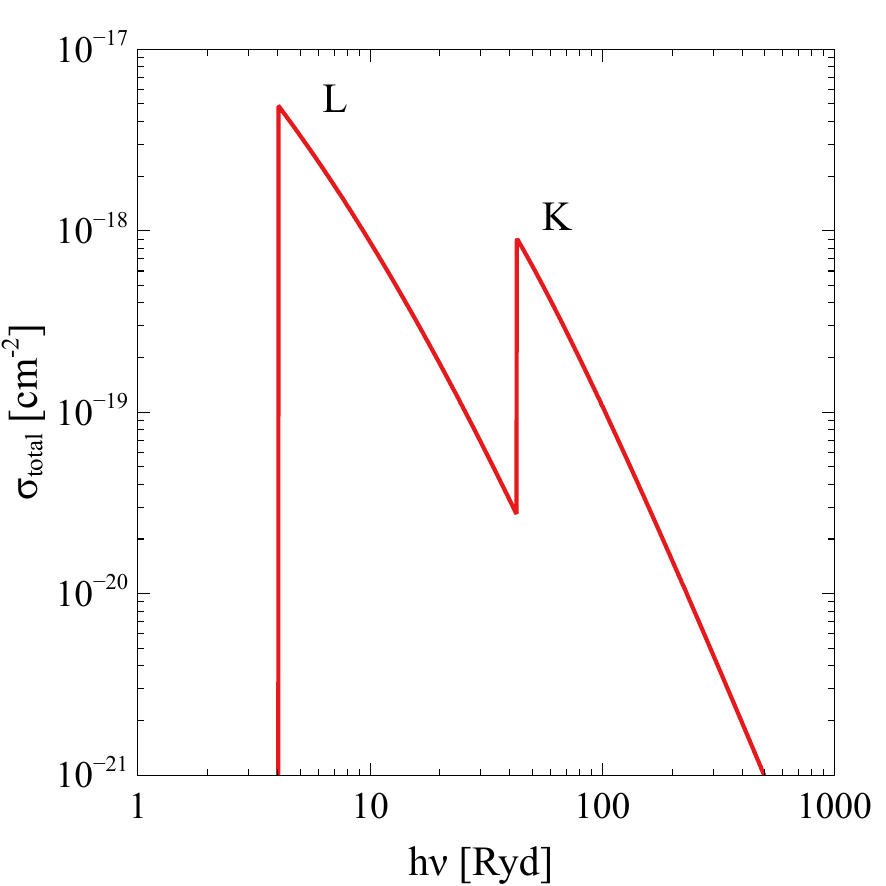}
\caption{
The total opacity of doubly ionized oxygen is shown.  The K and L shell edges are marked.
\label{fig:opacity}
}
\end{figure}

Consider the case of O$^{2+}$, a common ion of the 3$^{rd}$ most abundance element 
which produces the very strong O III lines.  
Three subshells, $1s^2$, $2s^2$, and $2p^2$, contribute to the total photoionization rate.  
We use the data fitted by \citep{1996ApJ...465..487V}.  
The cross sections are shown in Figure \ref{fig:opacity}.  
The K-shell cross sections are nearly one dex smaller than the valence shell.

\begin{figure}
\centering
\includegraphics[width=\columnwidth]{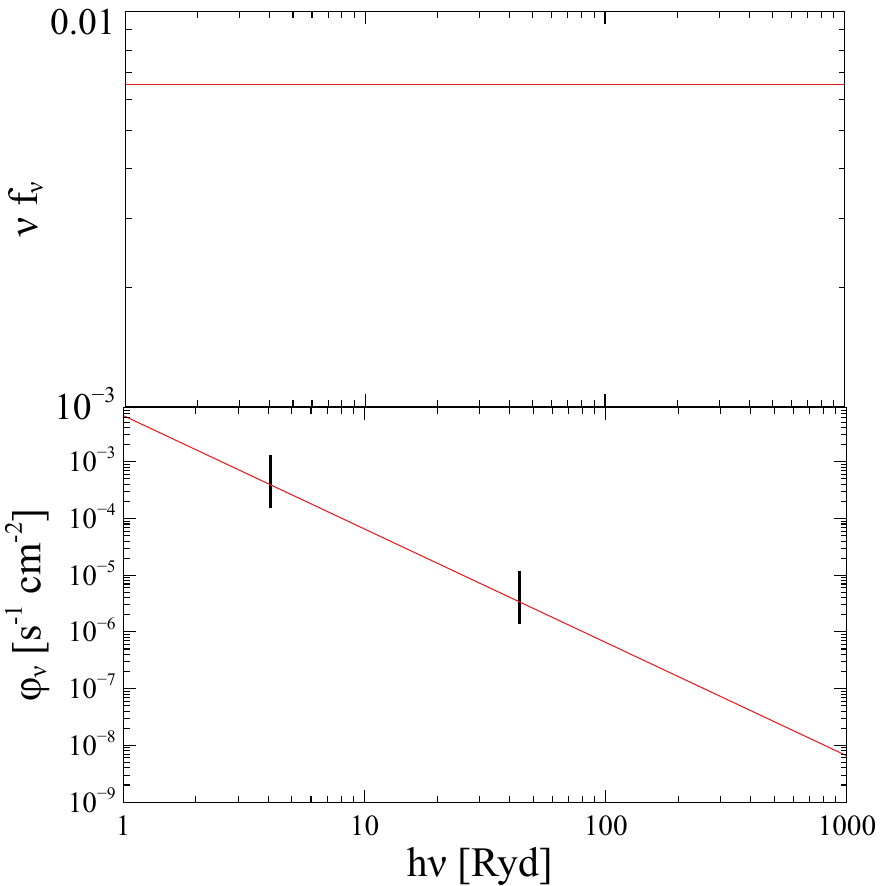}
\caption{
\label{fig:SED}
The upper panel shows the SED for an $f_\nu \propto \nu^{-1}$ SED,
an exceptionally hard continuum.
The lower panel gives the photon flux, 
$\phi_\nu \propto f_\nu / h\nu \propto \nu^{-2}$.
The hash marks indicate the locations of the $K$ and $L$ shells.
The photon flux in the $K$ shell is $\sim 2$ dex smaller than the $L$ shell.
}
\end{figure}

The radiation field shape enters in Equation \ref{eq:Gamma}. 
 Figure \ref{fig:SED} shows a power-law continuum, one with $f_\nu \propto \nu^{-1}$.  
 This is an exceptionally hard SED with a large number of K-shell photons compared with L-shell.  
Even quasars, with their non-thermal continuum, are not this hard.  
This will overestimate the importance of K-shell photoionization.
The upper panel of Figure 2 shows this SED as $\nu f_\nu$. 
 
The photon flux $\phi_\nu$ enters in the photoionization rate.  
This is the ratio $\phi_\nu \propto f_\nu / h\nu \propto \nu^{-2}$.  
This is shown in the lower panel.  
The flux of $K$-shell photons is about 150 times smaller than the flux of $L$ shell photons.

Cloudy includes a command to report each shell's photoionization rate $\Gamma_n$. 
We need this to treat Auger electron ejection and fluorescent emission properly.   
That rate is shown in the following table:

\begin{table}
\centering
\caption{\label{tab:PhotoRates}  Photoionization rates per subshell}
\begin{tabular}{ c c  }
\hline
Shell & $\Gamma_n$ \\
\hline
$1s^2$  & 8.84\e{-13}\\
$2s^2$  & 4.71\e{-14}\\
$2p^2$  & 1.28\e{-10} \\
\hline
\end{tabular}
\end{table}

The $L$ shell rates are about 150 times larger than the $K$ shell rates, 
due to the different photon fluxes, cross-sections, 
and energy ranges.  
The implication is that the $K$-shell rates are negligible compared with the valence shell rates.
One result is that differences in the X-ray portion of a SED will have little impact on
UV - IR emission. 

There were several papers published in the 1970s that discussed 
inner shell processes at length.  
This physics is fully included in Cloudy.  
Many of those papers overstated the importance of inner shell physics 
on the ionization.   

Although the inner shell physics has little effect on the ionization of the gas, 
it will be important for producing X-ray fluorescent emission lines as well as X-ray transmission spectroscopy.  
High-resolution X-ray observations of absorbing clouds in front of the 
X-ray continuum source could measure the features.

\section{Miscellaneous improvements}

The photoionization thresholds of most ions have been
slightly updated as of C17.02.
In this release, the entries for the radiative recombination continua on the
line stack have updated wavelengths to make them consistent with the new thresholds.

The ionization potentials employed by \Cloudy{} have been updated.
The new values are extracted from NIST and are current as of 2022-11-15.
The differences between the datasets for the elements up to Zinc are at most 0.5\%.
In addition, in preparation for future development, our ionization potential
dataset includes all natural elements and all their ions.
That is, our list extends up to Plutonium, instead of up to Krypton.

The solar abundances of \citet{Lodders2009} have been added to \Cloudy{}
as of C17.01.

To fix a bug where many blends were not predicted on the emergent
line stack, the requirements for adding lines to a blend have been tightened.
Blend components now have to be transmitted lines (i.e. have type 't', as
indicated in the \cdCommand{save line labels} output) that are associated to a
database transition. As a result, many blend components have been removed,
mostly predictions for the recombination contribution to a given line. These
predictions are based on ad hoc theories that are invalid over the entire
parameter range that \Cloudy\ covers. The components that have been removed
are still available as separate entries on the line stack. Detailed
information about the changes can be found on our wiki.\footnote{
\url{https://gitlab.nublado.org/cloudy/cloudy/-/wikis/NewC22}}

The label for the inward component of continuum bands has been changed to
``InwdBnd''. Previously that was ``Inwd''. This led to ambiguities with individual
\Ion{Fe}{II} lines when the big \Ion{Fe}{II} model was used.

The labels for collisional heating and cooling of the gas by grains have been
renamed to ``GrCH'' and ``GrCC'', respectively. Previously, they were both called
``GrGC''.

The code now uses the 2018 CODATA adjustment for the fundamental physical
constants.

\section{SEDs in \Cloudy}

Support for the \citet{KhaireSrianand2019} synthesis models of the
extragalactic background light has been added since C17.01. These SEDs cover
the range from the far infrared to TeV $\gamma$-rays, with an emphasis on the
extreme ultraviolet background responsible for the observed ionization of the
intergalactic medium for redshifts between $0 \leq z \leq 15$.

\begin{figure}
    \centering
    \includegraphics[width=\columnwidth]{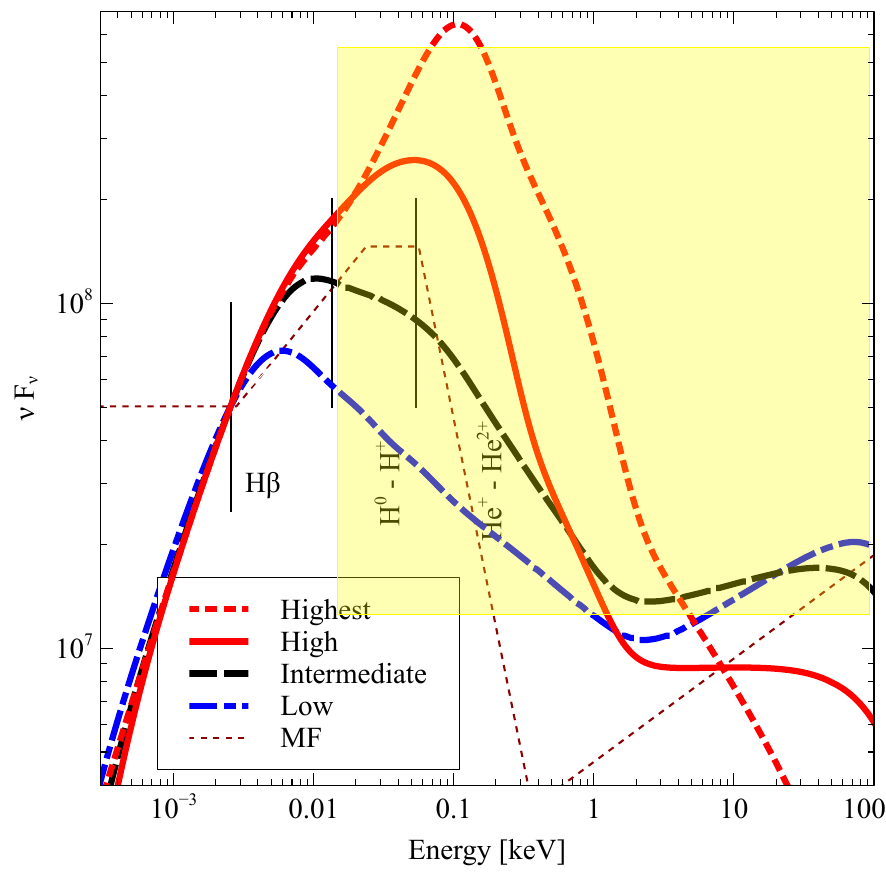}
    \caption{
        \label{fig:SEDsJin}
        This figure, from \citet{2020MNRAS.494.5917F},
        compares the SEDs of \citet{2012MNRAS.425..907J} and
        \citet{2017MNRAS.471..706J} over a range of Eddington ratio.
        The curve marked ``MF'' is taken from \citet{MathewsFerland87}
        and obtained with the \cdCommand{TABLE AGN} command.
    }
\end{figure}

In addition, four AGN SEDs described by \citet{2012MNRAS.425..907J}
and \citet{2017MNRAS.471..706J} are now included, see Fig.~\ref{fig:SEDsJin},
as well as the SED for NGC~5548 of \citet{2019ApJ...877..119D}.

\section{Infrastructure Changes}
\label{sec:infra}

\subsection{Migration to C++11}

The code has been ported to C++11. This version of the C++ standard delivers
significant new functionality and fixes many issues with the old C++98
standard we were using until now. Adopting the new standard enables writing
more versatile code. Compilers with full support for the C++11 standard have
been available since April 2013 (though some vendors only finalized their
support later). By now, these compilers should be widely available, and this
change should not impact the ability of our users to use \Cloudy. Users now
need a compiler with the following minimum requirements. For GNU g++ version
4.8.1 or later, for LLVM clang++ version 3.3 or later, for the Intel compiler
version 15.0 or later, and for the Oracle Developer Studio compiler version
12.5 or later.

\subsection{Executing the Code}

\Cloudy\ now supports the {\tt -e} flag on the command line. After this flag,
the user may enter one or more \Cloudy\ commands separated by a semicolon.
This averts the need to create an input file for very simple (test) models. It
can also be useful when calling \Cloudy\ from a shell script.

The executable now accepts the {\tt -s [ seed ]} command line flag. This will
set a fixed seed for the random number generator and is intended for debugging
and testing purposes. The parameter is a 64-bit seed in hexadecimal form. If
the seed is omitted, a default fixed value will be used. It is not recommended
to use this flag in normal \Cloudy\ runs.

The {\tt -a} flag on the command line has been removed. This flag was only
used in debugging and had been deprecated for some time.

The code has been enhanced to catch floating point exceptions and segmentation
faults on supported platforms (this includes all major compilers on Linux and
MacOS). This makes grid runs even more resilient against errors in one of the
grid models, allowing the code to finish the grid despite these errors.

If some models in a grid fail, the \cdCommand{save} output for that grid point
may be missing. As a workaround, the code now creates a stub file for the
missing output by taking the correct output from another grid point and
replacing all numbers by zeros.

In previous versions, when \Cloudy\ aborted (e.g. due to too many convergence
failures) it would try to soldier on and produce some (most likely wrong)
output despite the failure. This is no longer done. The code now stops
immediately after the abort.

Most compilers will now generate a backtrace of the call stack at the end of
the output if an error occurs. This is useful for debugging purposes.

\subsection{Reading and Writing Files}

For several years \Cloudy\ allowed the user to supply a custom search path for
searching data files. The development of this feature has now been finalized.
The code now uses a consistent policy for finding files that it needs to read.
All commands will look in the local directory first, and then in the data
directory (this is the standard search path). The user can alter the search
path by defining the environment variable CLOUDY\_DATA\_PATH before the code
is started up. This gives the user more freedom to choose where custom data
files are stored. Setting this variable will no longer affect the compilation of
the code (as was the case in C17 and before). The code will always write
output in the local directory. This is now enforced.

\subsection{Parser changes}

We have started modifying the parser to check the script's syntax more strictly. The long-term goal is to check everything that is typed.
In this release, we will start by fully checking the command name itself.
Abbreviating the command is still allowed (as was already the case in previous
releases) but all the typed characters will now be checked. We will
enforce US spelling where relevant.

The input deck's set of allowed comment characters has been significantly
reduced. Since version C17.01, it is no longer allowed to use ``c'' or ``C''
to start a comment. In the current version, only comments starting with ``\#''
or ``\#\#'' are allowed. Comments of the first type will be echoed in the main
output, while comments starting with ``\#\#'' will not be echoed.

The parser now supports line disambiguation. The \Cloudy\ line stack may
contain seemingly duplicate entries with the same label and wavelength but
which are actually different lines. This can create problems if you want to
use such a line. One example is the \Ion{H}{I} 4.65247~$\mu$m line, which may
be either the $7 \rightarrow 5$ or the $35 \rightarrow 7$ line. You can now
optionally supply the lower and upper-level index, or the energy of the lower
level, to indicate which line you want. The \cdCommand{save line labels}
output contains the necessary information to do this. This new syntax is
supported by all commands that read line identifications. It is also supported
by the subroutines \cdRoutine{cdLine}, \cdRoutine{cdEmis}, and
\cdRoutine{cdGetLineList}. This type of disambiguation is not possible for all
lines on the line stack.

\subsection{Numerical methods}

The old random number generators have been removed from \Cloudy. These were
based on the Mersenne twister algorithm and the Box-Muller method for
generating random numbers with a normal (Gaussian) distribution.
The new code uses a fully vectorized version of xoshiro256**
\citep{BlackmanVigna2021}\footnote{\url{http://xoshiro.di.unimi.it}.},
while the random numbers with a normal distribution are now generated
using the Ziggurat algorithm \citep{MarsagliaTsang00}.
Both methods are much faster than the old ones. An
additional advantage is that the new code is fully aware of parallelization in
the code, meaning that parallel ranks created with MPI or fork will
automatically have a different sequence of random numbers. The code now
generates a random seed at the start of execution by default (when available
derived from {\tt /dev/urandom}, otherwise using the system time), unless the
{\tt -s} command line parameter described above is used.

\subsection{New, modified, and deleted commands}

Since C17.01 the \cdCommand{save xspec} command has a new option
\cdCommand{normalize}. In versions prior to C17, the spectra would always have
the same normalization as the \cdCommand{save continuum} output. This can be
inconvenient for comparing spectra in grids where the normalization of the
spectra can be vastly different. When using the \cdCommand{normalize} keyword,
all spectra will be normalized to 1~photon\,cm$^{-2}$\,s$^{-1}$\,keV$^{-1}$ at
a photon energy of 1~keV. The user can alter the latter value to a different
photon energy.

Since C17.02, the \cdCommand{save transmitted continuum} command also works in
luminosity mode, and the keyword \cdCommand{last} is implicitly assumed to
avoid useless output. The format of the \cdCommand{save transmitted continuum}
file has changed, so files from versions C17.01 and older will no longer be
accepted. Spherical dilution will now be implicitly handled when the keyword
\cdCommand{scale} is used, and the first and second models both set a radius.

The command \cdCommand{database h-like keep fine structure} has been added.
This allows the fine-structure components of the hydrogen-like lines to be
reported on the line stack. Previous versions of the code already computed
these components but did not report them. This behavior is still the default,
but by including this command, the fine-structure lines will be added to the
line stack.

The commands \cdCommand{crash segfault}, \cdCommand{crash abort},
\cdCommand{crash grid}, and \cdCommand{crash bounds array} have been added to
emulate additional sources of errors. The \cdCommand{crash bounds heap}
command has been removed as \Cloudy{} no longer uses this method of allocating
memory. The \cdCommand{crash undefined} commands have been reorganized and the
only option left (without any additional keywords) is to test access to an
undefined variable on the stack (this used to be called \cdCommand{crash
  undefined stack \OR{} auto}).

The \cdCommand{set assert abort} command has been removed. Its effect was
identical to the {\tt -a} command line flag, which was only used for debugging
the code. The \cdCommand{stop nTotalIoniz} command has been removed. This was
a debugging tool that was very rarely used. The \cdCommand{drive} family of
commands have been removed. These were designed to test certain
aspects of the code interactively. They have not been used in a long time and are obsolete.
The \cdCommand{state} command has been removed. This was an unfinished
experimental feature to save or restore the state of the code. This project
has now been abandoned as it was too difficult to do. The \cdCommand{plot}
family of commands has been removed (as was already announced in \Hazy). This
was obsolete code for producing ASCII plots on line printers.

The option to set the seed for the random number generator has been removed
from the \cdCommand{database H2 noise} and \cdCommand{database H-like \OR{}
    He-like error generation} commands. Using a random seed is now the
default, so the user is no longer needed to set the seed.

The \cdCommand{table SED} command now accepts the \cdCommand{Flambda} keyword
in the SED data file. Fluxes in $F_\nu$ or $\nu F_\nu$ units were already
accepted, now $F_\lambda$ units can also be used.

The upper limit to the number of lines that can be supplied to the
\cdCommand{print line sum} and \cdCommand{save lines emissivity} commands has
been removed.

The command \cdCommand{set blend} has been added, allowing the user to define
custom line blends. This command allowed us to move most of the blends that
used to be hardwired into the code (e.g., Blnd 1909) to a new init file called
\cdFilename{blends.ini} that is part of the data directory. This file is
parsed automatically when starting up \Cloudy, unless the \cdCommand{no
  blends} command is included in the script.

The \cdCommand{abundances} command has been modified. It is now mandatory to
include an element symbol in front of the abundance, which will determine
what element the abundance belongs to. This makes the command much safer. It
also allows the user to put the elements in arbitrary order and removes the
need to complete the list of abundances. The \cdCommand{elements read}
command has been removed as it is no longer needed.

The \cdCommand{database He-like FSM} command has been removed. It was not
working correctly, and moreover, \citet{Bauman2005} found that as a result of
the principle of spectroscopic stability, it had very little impact on the
predictions for the He\,{\sc i} spectrum. The situation is different though
for highly charged ions, e.g. Fe\,{\sc xxv}. For such ions, the individual
fine-structure components can be spectroscopically resolved and treating
fine-structure mixing would be warranted. Implementing this will be postponed
to a future release.

\subsection{Storing SED Grids}

The SED grids supported by \Cloudy\ no longer need to be compiled into binary
form. The code now directly reads the ASCII files to obtain the necessary
information. Compiling the ASCII files is still supported, but now produces a
completely different type of file that contains indices into the ASCII file.
This step is optional but is strongly recommended for large grids to speed up
the code. With this setup, recompiling stellar atmosphere grids is no longer necessary when the frequency mesh is changed. Compiling SEDs in an
external format (such as Starburst99 or the Rauch stellar atmosphere grids)
is still mandatory to obtain the ASCII files.

\section{Future Directions}
\label{sec:future}

Work is under way to extend and improve \Cloudy{}.
Some of these features will be available, at least in part, by the time
of the next release, while others may require longer to come to fruition.
Some of these directions were already discussed by \citet{vanHoof2020}.

As explained above, the X-ray capabilities of \Cloudy{} have been extended
substantially.
Yet, more work remains to be done.
We are currently working on resolving the doublets of Lyman-like emission lines
($np \to 1s$ transitions) in hydrogenic ions (Gunasekera, in preparation).
This feature should be available in the next major release.
In addition, we plan to extend \Cloudy{} to include the results of experiments on
inner-shell ionization.
In the longer run, we should update the charge-exchange data of the code with
modern calculations, e.g., with the \texttt{Kronos}
database\footnote{\url{https://www.physast.uga.edu/research/stancil-group/atomic-molecular-databases/kronos}}
\citep[e.g.,][]{Mullen2016-CX-FeXXV,Cumbee2016-CX,Cumbee2018-CX,Lyons2017-CX}.

Over the last one or two decades, astronomy has entered its high-precision era.
So must \Cloudy{}.
Given that much of what we know about the chemical composition and kinematics
of celestial sources comes from spectroscopy, it is of paramount importance to
improve the atomic data the code employs to make quantitative predictions for,
and to interpret observations.
A program is underway (PI: Chatzikos) to produce high-quality atomic data that combine
laboratory-grade wavelengths (i.e., energies) with accurate transition probabilities
and collision strengths.
The new models will be added to our Stout database.

Other aspects of \Cloudy{} that are under active development include
time-dependent calculations and the radiative transfer module of the code.
We aim to publicly release these updates in the next one or two releases.

\acknowledgments

MC acknowledges support from NSF (1910687), NASA (19-ATP19-0188, 22-ADAP22-0139),
and STScI (HST-AR-14556.001-A).
GJF acknowledges support by NSF (1816537, 1910687), NASA (ATP 17-ATP17-0141, 19-ATP19-0188), and STScI (HST-AR- 15018 and HST-GO-16196.003-A).
SB acknowledges support from
PRIN MUR 2017 ``Black hole winds and the baryon life cycle of galaxies: the stone-guest at the galaxy evolution supper'' and from the European Union Horizon 2020 Research and Innovation Framework Programme under grant agreement AHEAD2020 n.871158.
GS acknowledges WOS-A grant from the Department of Science and Technology (SR/WOS-A/PM-2/2021).
GS thanks Ziwei Zhang for providing the vib-rotational data of SiS.

\bibliography{Cloudy23} 

\begin{thebibliography}
\expandafter\ifx\csname natexlab\endcsname\relax\def\natexlab#1{#1}\fi
\expandafter\ifx\csname href\endcsname\relax
  \def\href#1#2{}\fi
\expandafter\ifx\csname urllinklabel\endcsname\relax
  \def\urllinklabel{[LINK]}\fi
\expandafter\ifx\csname adsurllinklabel\endcsname\relax
  \def\adsurllinklabel{[ADS]}\fi

\bibitem[{{Amato} {et~al.}(2021){Amato}, {Grinberg}, {Hell}, {Bianchi},
  {Pinto}, {D'A{\'\i}}, {Del Santo}, {Mineo}, \&
  {Santangelo}}]{2021A&A...648A.105A}
{Amato}, R., {Grinberg}, V., {Hell}, N., {Bianchi}, S., {Pinto}, C.,
  {D'A{\'\i}}, A., {Del Santo}, M., {Mineo}, T., \& {Santangelo}. 2021, \aap,
  648, A105


\bibitem[{Anusuri(2019)}]{ANUSURI2019e01647}
Anusuri, B. 2019, Heliyon, 5, e01647
 \href{https://www.sciencedirect.com/science/article/pii/S2405844019316330}{\urllinklabel}

\bibitem[{{Badnell} \& {Ballance}(2014)}]{BadnellBalance14}
{Badnell}, N.~R. \& {Ballance}, C.~P. 2014, \apj, 785, 99


\bibitem[{{Badnell} {et~al.}(2021){Badnell}, {Guzm{\'a}n}, {Brodie},
  {Williams}, {van Hoof}, {Chatzikos}, \& {Ferland}}]{Badnell-lchanging}
{Badnell}, N.~R., {Guzm{\'a}n}, F., {Brodie}, S., {Williams}, R.~J.~R., {van
  Hoof}, P.~A.~M., {Chatzikos}, M., \& {Ferland}, G.~J. 2021, \mnras, 507, 2922


\bibitem[{{Baker} \& {Menzel}(1938)}]{1938ApJ....88...52B}
{Baker}, J.~G. \& {Menzel}, D.~H. 1938, \apj, 88, 52


\bibitem[{{Baldwin} {et~al.}(2004){Baldwin}, {Ferland}, {Korista}, {Hamann}, \&
  {LaCluyz{\'e}}}]{2004ApJ...615..610B}
{Baldwin}, J.~A., {Ferland}, G.~J., {Korista}, K.~T., {Hamann}, F., \&
  {LaCluyz{\'e}}, A. 2004, \apj, 615, 610


\bibitem[{{Bauman} {et~al.}(2005){Bauman}, {Porter}, {Ferland}, \&
  {MacAdam}}]{Bauman2005}
{Bauman}, R.~P., {Porter}, R.~L., {Ferland}, G.~J., \& {MacAdam}, K.~B. 2005,
  \apj, 628, 541


\bibitem[{{Bautista} {et~al.}(2015){Bautista}, {Fivet}, {Ballance}, {Quinet},
  {Ferland}, {Mendoza}, \& {Kallman}}]{2015ApJ...808..174B}
{Bautista}, M.~A., {Fivet}, V., {Ballance}, C., {Quinet}, P., {Ferland}, G.,
  {Mendoza}, C., \& {Kallman}, T.~R. 2015, \apj, 808, 174


\bibitem[{Blackman \& Vigna(2021)}]{BlackmanVigna2021}
Blackman, D. \& Vigna, S. 2021, ACM Trans. Math. Softw., 47
 \href{https://doi.org/10.1145/3460772}{\urllinklabel}

\bibitem[{{Bruhweiler} \& {Verner}(2008)}]{Bruhweiler_2008}
{Bruhweiler}, F. \& {Verner}, E. 2008, \apj, 675, 83


\bibitem[{{Bundy} {et~al.}(2015){Bundy}, {Bershady}, {Law}, {Yan}, {Drory},
  {MacDonald}, {Wake}, {Cherinka}, {S{\'a}nchez-Gallego}, {Weijmans}, {Thomas},
  {Tremonti}, {Masters}, {Coccato}, {Diamond-Stanic}, {Arag{\'o}n-Salamanca},
  {Avila-Reese}, {Badenes}, {Falc{\'o}n-Barroso}, {Belfiore}, {Bizyaev},
  {Blanc}, {Bland-Hawthorn}, {Blanton}, {Brownstein}, {Byler}, {Cappellari},
  {Conroy}, {Dutton}, {Emsellem}, {Etherington}, {Frinchaboy}, {Fu}, {Gunn},
  {Harding}, {Johnston}, {Kauffmann}, {Kinemuchi}, {Klaene}, {Knapen},
  {Leauthaud}, {Li}, {Lin}, {Maiolino}, {Malanushenko}, {Malanushenko}, {Mao},
  {Maraston}, {McDermid}, {Merrifield}, {Nichol}, {Oravetz}, {Pan}, {Parejko},
  {Sanchez}, {Schlegel}, {Simmons}, {Steele}, {Steinmetz}, {Thanjavur},
  {Thompson}, {Tinker}, {van den Bosch}, {Westfall}, {Wilkinson}, {Wright},
  {Xiao}, \& {Zhang}}]{bundy2015}
{Bundy}, K., {Bershady}, M.~A., {Law}, D.~R., {Yan}, R., {Drory}, N.,
  {MacDonald}, N., {Wake}, D.~A., {Cherinka}, B., {S{\'a}nchez-Gallego}, J.~R.,
  {Weijmans}, A.-M., {Thomas}, D., {Tremonti}, C., {Masters}, K., {Coccato},
  L., {Diamond-Stanic}, A.~M., {Arag{\'o}n-Salamanca}, A., {Avila-Reese}, V.,
  {Badenes}, C., {Falc{\'o}n-Barroso}, J., {Belfiore}, F., {Bizyaev}, D.,
  {Blanc}, G.~A., {Bland-Hawthorn}, J., {Blanton}, M.~R., {Brownstein}, J.~R.,
  {Byler}, N., {Cappellari}, M., {Conroy}, C., {Dutton}, A.~A., {Emsellem}, E.,
  {Etherington}, J., {Frinchaboy}, P.~M., {Fu}, H., {Gunn}, J.~E., {Harding},
  P., {Johnston}, E.~J., {Kauffmann}, G., {Kinemuchi}, K., {Klaene}, M.~A.,
  {Knapen}, J.~H., {Leauthaud}, A., {Li}, C., {Lin}, L., {Maiolino}, R.,
  {Malanushenko}, V., {Malanushenko}, E., {Mao}, S., {Maraston}, C.,
  {McDermid}, R.~M., {Merrifield}, M.~R., {Nichol}, R.~C., {Oravetz}, D.,
  {Pan}, K., {Parejko}, J.~K., {Sanchez}, S.~F., {Schlegel}, D., {Simmons}, A.,
  {Steele}, O., {Steinmetz}, M., {Thanjavur}, K., {Thompson}, B.~A., {Tinker},
  J.~L., {van den Bosch}, R.~C.~E., {Westfall}, K.~B., {Wilkinson}, D.,
  {Wright}, S., {Xiao}, T., \& {Zhang}, K. 2015, \apj, 798, 7


\bibitem[{{Camilloni} {et~al.}(2021){Camilloni}, {Bianchi}, {Amato}, {Ferland},
  \& {Grinberg}}]{2021RNAAS...5..149C}
{Camilloni}, F., {Bianchi}, S., {Amato}, R., {Ferland}, G., \& {Grinberg}, V.
  2021, Research Notes of the American Astronomical Society, 5, 149


\bibitem[{{Chakraborty} {et~al.}(2020{\natexlab{a}}){Chakraborty}, {Ferland},
  {Bianchi}, \& {Chatzikos}}]{KaCorr}
{Chakraborty}, P., {Ferland}, G.~J., {Bianchi}, S., \& {Chatzikos}, M.
  2020{\natexlab{a}}, Research Notes of the American Astronomical Society, 4,
  184


\bibitem[{{Chakraborty} {et~al.}(2022){Chakraborty}, {Ferland}, {Chatzikos},
  {Fabian}, {Bianchi}, {Guzm{\'a}n}, \& {Su}}]{Chakraborty.IV}
{Chakraborty}, P., {Ferland}, G.~J., {Chatzikos}, M., {Fabian}, A.~C.,
  {Bianchi}, S., {Guzm{\'a}n}, F., \& {Su}, Y. 2022, \apj, 935, 70


\bibitem[{{Chakraborty} {et~al.}(2020{\natexlab{b}}){Chakraborty}, {Ferland},
  {Chatzikos}, {Guzm{\'a}n}, \& {Su}}]{Chakraborty.I}
{Chakraborty}, P., {Ferland}, G.~J., {Chatzikos}, M., {Guzm{\'a}n}, F., \&
  {Su}, Y. 2020{\natexlab{b}}, \apj, 901, 68


\bibitem[{{Chakraborty} {et~al.}(2020{\natexlab{c}}){Chakraborty}, {Ferland},
  {Chatzikos}, {Guzm{\'a}n}, \& {Su}}]{Chakraborty.II}
---. 2020{\natexlab{c}}, \apj, 901, 69


\bibitem[{{Chakraborty} {et~al.}(2021){Chakraborty}, {Ferland}, {Chatzikos},
  {Guzm{\'a}n}, \& {Su}}]{Chakraborty.III}
---. 2021, \apj, 912, 26


\bibitem[{{Chamberlain}(1953)}]{1953ApJ...117..399C}
{Chamberlain}, J.~W. 1953, \apj, 117, 399


\bibitem[{{Cumbee} {et~al.}(2016){Cumbee}, {Liu}, {Lyons}, {Schultz},
  {Stancil}, {Wang}, \& {Ali}}]{Cumbee2016-CX}
{Cumbee}, R.~S., {Liu}, L., {Lyons}, D., {Schultz}, D.~R., {Stancil}, P.~C.,
  {Wang}, J.~G., \& {Ali}, R. 2016, \mnras, 458, 3554


\bibitem[{{Cumbee} {et~al.}(2018){Cumbee}, {Mullen}, {Lyons}, {Shelton},
  {Fogle}, {Schultz}, \& {Stancil}}]{Cumbee2018-CX}
{Cumbee}, R.~S., {Mullen}, P.~D., {Lyons}, D., {Shelton}, R.~L., {Fogle}, M.,
  {Schultz}, D.~R., \& {Stancil}, P.~C. 2018, \apj, 852, 7


\bibitem[{{Dehghanian} {et~al.}(2019){Dehghanian}, {Ferland}, {Kriss},
  {Peterson}, {Mathur}, {Mehdipour}, {Guzm{\'a}n}, {Chatzikos}, {van Hoof},
  {Williams}, {Arav}, {Barth}, {Bentz}, {Bisogni}, {Brandt}, {Crenshaw}, {Dalla
  Bont{\`a}}, {De Rosa}, {Fausnaugh}, {Gelbord}, {Goad}, {Gupta}, {Horne},
  {Kaastra}, {Knigge}, {Korista}, {McHardy}, {Pogge}, {Starkey}, \&
  {Vestergaard}}]{2019ApJ...877..119D}
{Dehghanian}, M., {Ferland}, G.~J., {Kriss}, G.~A., {Peterson}, B.~M.,
  {Mathur}, S., {Mehdipour}, M., {Guzm{\'a}n}, F., {Chatzikos}, M., {van Hoof},
  P.~A.~M., {Williams}, R.~J.~R., {Arav}, N., {Barth}, A.~J., {Bentz}, M.~C.,
  {Bisogni}, S., {Brandt}, W.~N., {Crenshaw}, D.~M., {Dalla Bont{\`a}}, E., {De
  Rosa}, G., {Fausnaugh}, M.~M., {Gelbord}, J.~M., {Goad}, M.~R., {Gupta}, A.,
  {Horne}, K., {Kaastra}, J., {Knigge}, C., {Korista}, K.~T., {McHardy}, I.~M.,
  {Pogge}, R.~W., {Starkey}, D.~A., \& {Vestergaard}, M. 2019, \apj, 877, 119


\bibitem[{{Doddipatla} {et~al.}(2021){Doddipatla}, {He}, {Goettl}, {Kaiser},
  {Galv{\~a}o}, \& {Millar}}]{2021SciA....7.7003D}
{Doddipatla}, S., {He}, C., {Goettl}, S.~J., {Kaiser}, R.~I., {Galv{\~a}o}, B.
  R.~L., \& {Millar}, T.~J. 2021, Science Advances, 7, eabg7003


\bibitem[{{Draine}(2003)}]{Draine2003}
{Draine}, B.~T. 2003, \apj, 598, 1026


\bibitem[{{Ferland}(1999)}]{1999PASP..111.1524F}
{Ferland}, G.~J. 1999, \pasp, 111, 1524


\bibitem[{{Ferland} {et~al.}(2017){Ferland}, {Chatzikos}, {Guzm{\'a}n},
  {Lykins}, {van Hoof}, {Williams}, {Abel}, {Badnell}, {Keenan}, {Porter}, \&
  {Stancil}}]{2017RMxAA..53..385F}
{Ferland}, G.~J., {Chatzikos}, M., {Guzm{\'a}n}, F., {Lykins}, M.~L., {van
  Hoof}, P.~A.~M., {Williams}, R.~J.~R., {Abel}, N.~P., {Badnell}, N.~R.,
  {Keenan}, F.~P., {Porter}, R.~L., \& {Stancil}, P.~C. 2017, Revista Mexicana
  de Astronomia y Astrofisica, 53, 385


\bibitem[{{Ferland} {et~al.}(2020){Ferland}, {Done}, {Jin}, {Landt}, \&
  {Ward}}]{2020MNRAS.494.5917F}
{Ferland}, G.~J., {Done}, C., {Jin}, C., {Landt}, H., \& {Ward}, M.~J. 2020,
  \mnras, 494, 5917


\bibitem[{{Ferland} {et~al.}(1998){Ferland}, {Korista}, {Verner}, {Ferguson},
  {Kingdon}, \& {Verner}}]{1998PASP..110..761F}
{Ferland}, G.~J., {Korista}, K.~T., {Verner}, D.~A., {Ferguson}, J.~W.,
  {Kingdon}, J.~B., \& {Verner}, E.~M. 1998, \pasp, 110, 761


\bibitem[{{Ferland} {et~al.}(2013){Ferland}, {Porter}, {van Hoof}, {Williams},
  {Abel}, {Lykins}, {Shaw}, {Henney}, \& {Stancil}}]{2013RMxAA..49..137F}
{Ferland}, G.~J., {Porter}, R.~L., {van Hoof}, P.~A.~M., {Williams}, R.~J.~R.,
  {Abel}, N.~P., {Lykins}, M.~L., {Shaw}, G., {Henney}, W.~J., \& {Stancil},
  P.~C. 2013, Revista Mexicana de Astronomia y Astrofisica, 49, 137


\bibitem[{{Fern{\'a}ndez-Menchero} {et~al.}(2014){Fern{\'a}ndez-Menchero}, {Del
  Zanna}, \& {Badnell}}]{2014A&A...572A.115F}
{Fern{\'a}ndez-Menchero}, L., {Del Zanna}, G., \& {Badnell}, N.~R. 2014, \aap,
  572, A115


\bibitem[{Fujimoto(1978)}]{Fujimoto1978}
Fujimoto, T. 1978, Semi-empirical Cross Sections and Rate Coefficients for
  Excitation and Ionization by Electron Collision and Photoionization of
  Helium, Tech. Rep. IPPJ-AM-8, Nagoya University


\bibitem[{{Gunasekera} {et~al.}(2022{\natexlab{a}}){Gunasekera}, {Chatzikos},
  \& {Ferland}}]{CloudyChianti10}
{Gunasekera}, C.~M., {Chatzikos}, M., \& {Ferland}, G.~J. 2022{\natexlab{a}},
  Astronomy, 1, 255


\bibitem[{{Gunasekera} {et~al.}(2022{\natexlab{b}}){Gunasekera}, {Ji},
  {Chatzikos}, {Yan}, \& {Ferland}}]{CloudyGrainDepl}
{Gunasekera}, C.~M., {Ji}, X., {Chatzikos}, M., {Yan}, R., \& {Ferland}, G.
  2022{\natexlab{b}}, \mnras, 512, 2310


\bibitem[{{Gunasekera} {et~al.}(2023){Gunasekera}, {Ji}, {Chatzikos}, {Yan}, \&
  {Ferland}}]{Gunasekera2023}
---. 2023, \mnras, 520, 4345


\bibitem[{{Guzm{\'a}n} {et~al.}(2017{\natexlab{a}}){Guzm{\'a}n}, {Badnell},
  {Chatzikos}, {van Hoof}, {Williams}, \& {Ferland}}]{Guzman.2017.TwoPhoton}
{Guzm{\'a}n}, F., {Badnell}, N.~R., {Chatzikos}, M., {van Hoof}, P.~A.~M.,
  {Williams}, R.~J.~R., \& {Ferland}, G.~J. 2017{\natexlab{a}}, \mnras, 467,
  3944


\bibitem[{{Guzm{\'a}n} {et~al.}(2016){Guzm{\'a}n}, {Badnell}, {Williams}, {van
  Hoof}, {Chatzikos}, \& {Ferland}}]{Guzman.I.2016}
{Guzm{\'a}n}, F., {Badnell}, N.~R., {Williams}, R.~J.~R., {van Hoof}, P.~A.~M.,
  {Chatzikos}, M., \& {Ferland}, G.~J. 2016, \mnras, 459, 3498


\bibitem[{{Guzm{\'a}n} {et~al.}(2017{\natexlab{b}}){Guzm{\'a}n}, {Badnell},
  {Williams}, {van Hoof}, {Chatzikos}, \& {Ferland}}]{Guzman.II.2017}
---. 2017{\natexlab{b}}, \mnras, 464, 312


\bibitem[{{Guzm{\'a}n} {et~al.}(2019){Guzm{\'a}n}, {Chatzikos}, {van Hoof},
  {Balser}, {Dehghanian}, {Badnell}, \& {Ferland }}]{Guzman.III.2019}
{Guzm{\'a}n}, F., {Chatzikos}, M., {van Hoof}, P.~A.~M., {Balser}, D.~S.,
  {Dehghanian}, M., {Badnell}, N.~R., \& {Ferland }, G.~J. 2019, \mnras, 486,
  1003


\bibitem[{{Hell} {et~al.}(2016){Hell}, {Brown}, {Wilms}, {Grinberg},
  {Clementson}, {Liedahl}, {Porter}, {Kelley}, {Kilbourne}, \&
  {Beiersdorfer}}]{2016ApJ...830...26H}
{Hell}, N., {Brown}, G.~V., {Wilms}, J., {Grinberg}, V., {Clementson}, J.,
  {Liedahl}, D., {Porter}, F.~S., {Kelley}, R.~L., {Kilbourne}, C.~A., \&
  {Beiersdorfer}, P. 2016, \apj, 830, 26


\bibitem[{{Jenkins}(2009)}]{Jenkins2009}
{Jenkins}, E.~B. 2009, \apj, 700, 1299


\bibitem[{{Jin} {et~al.}(2017){Jin}, {Done}, {Ward}, \&
  {Gardner}}]{2017MNRAS.471..706J}
{Jin}, C., {Done}, C., {Ward}, M., \& {Gardner}, E. 2017, Monthly Notices of
  the Royal Astronomical Society, 471, 706


\bibitem[{{Jin} {et~al.}(2012){Jin}, {Ward}, \& {Done}}]{2012MNRAS.425..907J}
{Jin}, C., {Ward}, M., \& {Done}, C. 2012, Monthly Notices of the Royal
  Astronomical Society, 425, 907


\bibitem[{{Kaastra} \& {Mewe}(1993)}]{1993A&AS...97..443K}
{Kaastra}, J.~S. \& {Mewe}, R. 1993, \aaps, 97, 443


\bibitem[{{Khaire} \& {Srianand}(2019)}]{KhaireSrianand2019}
{Khaire}, V. \& {Srianand}, R. 2019, \mnras, 484, 4174


\bibitem[{{Komasa} {et~al.}(2011){Komasa}, {Piszczatowski}, {\L ach},
  {Przybytek}, {Jeziorski}, \& {Pachucki}}]{Komasa16H2Eneries}
{Komasa}, J., {Piszczatowski}, K., {\L ach}, G., {Przybytek}, M., {Jeziorski},
  B., \& {Pachucki}, K. 2011, Journal of Chemical Theory and Computation, 7,
  3105, pMID: 26598154
 \href{http://dx.doi.org/10.1021/ct200438t}{\urllinklabel}

\bibitem[{{Kramida} {et~al.}(2018){Kramida}, {Ralchenko}, {Nave}, \&
  {Reader}}]{2018APS..DMPM01004K}
{Kramida}, A., {Ralchenko}, Y., {Nave}, G., \& {Reader}, J. 2018, in APS
  Meeting Abstracts, Vol. 2018, APS Division of Atomic, Molecular and Optical
  Physics Meeting Abstracts, M01.004


\bibitem[{{Laha} {et~al.}(2017){Laha}, {Tyndall}, {Keenan}, {Ballance},
  {Ramsbottom}, {Ferland}, \& {Hibbert}}]{2017ApJ...841....3L}
{Laha}, S., {Tyndall}, N.~B., {Keenan}, F.~P., {Ballance}, C.~P., {Ramsbottom},
  C.~A., {Ferland}, G.~J., \& {Hibbert}, A. 2017, \apj, 841, 3


\bibitem[{{Lebedev} \& {Beigman}(1998)}]{Lebedevandbeigman1998}
{Lebedev}, V.~S. \& {Beigman}, I.~L. 1998, {Physics of Highly Excited Atoms and
  Ions} (Springer)


\bibitem[{{Liang} \& {Badnell}(2011)}]{2011A&A...528A..69L}
{Liang}, G.~Y. \& {Badnell}, N.~R. 2011, \aap, 528, A69


\bibitem[{{Liang} {et~al.}(2009){Liang}, {Whiteford}, \&
  {Badnell}}]{2009A&A...500.1263L}
{Liang}, G.~Y., {Whiteford}, A.~D., \& {Badnell}, N.~R. 2009, \aap, 500, 1263


\bibitem[{{Liedahl}(2005)}]{2005AIPC..774...99L}
{Liedahl}, D.~A. 2005, in American Institute of Physics Conference Series, Vol.
  774, X-ray Diagnostics of Astrophysical Plasmas: Theory, Experiment, and
  Observation, ed. R.~{Smith}, 99--108


\bibitem[{{Lique}(2015)}]{Lique15}
{Lique}, F. 2015, \mnras, 453, 810


\bibitem[{{Lodders} {et~al.}(2009){Lodders}, {Palme}, \& {Gail}}]{Lodders2009}
{Lodders}, K., {Palme}, H., \& {Gail}, H.-P. 2009, Landolt B{\"o}rnstein, 44


\bibitem[{{Lykins} {et~al.}(2015){Lykins}, {Ferland}, {Kisielius}, {Chatzikos},
  {Porter}, {van Hoof}, {Williams}, {Keenan}, \& {Stancil}}]{Lykins2015}
{Lykins}, M.~L., {Ferland}, G.~J., {Kisielius}, R., {Chatzikos}, M., {Porter},
  R.~L., {van Hoof}, P.~A.~M., {Williams}, R.~J.~R., {Keenan}, F.~P., \&
  {Stancil}, P.~C. 2015, \apj, 807, 118


\bibitem[{{Lyons} {et~al.}(2017){Lyons}, {Cumbee}, \& {Stancil}}]{Lyons2017-CX}
{Lyons}, D., {Cumbee}, R.~S., \& {Stancil}, P.~C. 2017, \apjs, 232, 27


\bibitem[{{Marsaglia} \& {Tsang}(2000)}]{MarsagliaTsang00}
{Marsaglia}, G. \& {Tsang}, W.~W. 2000, Journal of Statistical Software, 5, 8


\bibitem[{{Mathews} \& {Ferland}(1987)}]{MathewsFerland87}
{Mathews}, W.~G. \& {Ferland}, G.~J. 1987, \apj, 323, 456


\bibitem[{{Matt} {et~al.}(1996){Matt}, {Brandt}, \&
  {Fabian}}]{1996MNRAS.280..823M}
{Matt}, G., {Brandt}, W.~N., \& {Fabian}, A.~C. 1996, \mnras, 280, 823


\bibitem[{{Menzel}(1937)}]{1937ApJ....85..330M}
{Menzel}, D.~H. 1937, \apj, 85, 330


\bibitem[{{Mullen} {et~al.}(2016){Mullen}, {Cumbee}, {Lyons}, \&
  {Stancil}}]{Mullen2016-CX-FeXXV}
{Mullen}, P.~D., {Cumbee}, R.~S., {Lyons}, D., \& {Stancil}, P.~C. 2016, \apjs,
  224, 31


\bibitem[{{Netzer}(2020)}]{Netzer2020}
{Netzer}, H. 2020, \mnras, 494, 1611


\bibitem[{{Osterbrock} \& {Ferland}(2006{\natexlab{a}})}]{AGN3}
{Osterbrock}, D.~E. \& {Ferland}, G.~J. 2006{\natexlab{a}}, {Astrophysics of
  gaseous nebulae and active galactic nuclei, 2nd.~ed.} (Sausalito, CA:
  University Science Books)


\bibitem[{{Osterbrock} \& {Ferland}(2006{\natexlab{b}})}]{2006agna.book.....O}
---. 2006{\natexlab{b}}, {Astrophysics of gaseous nebulae and active galactic
  nuclei, 2nd.~ed.} (Sausalito, CA: University Science Books)


\bibitem[{{Peimbert} {et~al.}(2017){Peimbert}, {Peimbert}, \&
  {Delgado-Inglada}}]{2017PASP..129h2001P}
{Peimbert}, M., {Peimbert}, A., \& {Delgado-Inglada}, G. 2017, \pasp, 129,
  082001


\bibitem[{{Pengelly} \& {Seaton}(1964)}]{PengellySeaton1964}
{Pengelly}, R.~M. \& {Seaton}, M.~J. 1964, \mnras, 127, 165


\bibitem[{{Priestley} {et~al.}(2017){Priestley}, {Barlow}, \&
  {Viti}}]{2017MNRAS.472.4444P}
{Priestley}, F.~D., {Barlow}, M.~J., \& {Viti}, S. 2017, \mnras, 472, 4444


\bibitem[{{R{\"o}llig}(2011)}]{2011A&A...530A...9R}
{R{\"o}llig}, M. 2011, \aap, 530, A9


\bibitem[{{Ross} {et~al.}(1978){Ross}, {Weaver}, \&
  {McCray}}]{1978ApJ...219..292R}
{Ross}, R.~R., {Weaver}, R., \& {McCray}, R. 1978, \apj, 219, 292


\bibitem[{{Roueff} {et~al.}(2014){Roueff}, {Alekseyev}, \& {Le
  Bourlot}}]{2014A&A...566A..30R}
{Roueff}, E., {Alekseyev}, A.~B., \& {Le Bourlot}, J. 2014, \aap, 566, A30


\bibitem[{{Sarkar} {et~al.}(2021){Sarkar}, {Ferland}, {Chatzikos},
  {Guzm{\'a}n}, {van Hoof}, {Smyth}, {Ramsbottom}, {Keenan}, \&
  {Ballance}}]{CloudyFeII2021}
{Sarkar}, A., {Ferland}, G.~J., {Chatzikos}, M., {Guzm{\'a}n}, F., {van Hoof},
  P.~A.~M., {Smyth}, R.~T., {Ramsbottom}, C.~A., {Keenan}, F.~P., \&
  {Ballance}, C.~P. 2021, \apj, 907, 12


\bibitem[{{Schilke} {et~al.}(2014){Schilke}, {Neufeld}, {M{\"u}ller}, {Comito},
  {Bergin}, {Lis}, {Gerin}, {Black}, {Wolfire}, {Indriolo}, {Pearson},
  {Menten}, {Winkel}, {S{\'a}nchez-Monge}, {M{\"o}ller}, {Godard}, \&
  {Falgarone}}]{2014A&A...566A..29S}
{Schilke}, P., {Neufeld}, D.~A., {M{\"u}ller}, H.~S.~P., {Comito}, C.,
  {Bergin}, E.~A., {Lis}, D.~C., {Gerin}, M., {Black}, J.~H., {Wolfire}, M.,
  {Indriolo}, N., {Pearson}, J.~C., {Menten}, K.~M., {Winkel}, B.,
  {S{\'a}nchez-Monge}, {\'A}., {M{\"o}ller}, T., {Godard}, B., \& {Falgarone},
  E. 2014, \aap, 566, A29


\bibitem[{{Shaw} {et~al.}(2023{\natexlab{a}}){Shaw}, {Ferland}, \&
  {Chatzikos}}]{2023RNAAS...7...45S}
{Shaw}, G., {Ferland}, G., \& {Chatzikos}, M. 2023{\natexlab{a}}, Research
  Notes of the American Astronomical Society, 7, 45


\bibitem[{{Shaw} {et~al.}(2023{\natexlab{b}}){Shaw}, {Ferland}, \&
  {Chatzikos}}]{2023RNAAS...7..153S}
---. 2023{\natexlab{b}}, Research Notes of the American Astronomical Society,
  7, 153


\bibitem[{{Shaw} \& {Ferland}(2021)}]{2021ApJ...908..138S}
{Shaw}, G. \& {Ferland}, G.~J. 2021, \apj, 908, 138


\bibitem[{{Shaw} {et~al.}(2005){Shaw}, {Ferland}, {Abel}, {Stancil}, \& {van
  Hoof}}]{Shaw2005}
{Shaw}, G., {Ferland}, G.~J., {Abel}, N.~P., {Stancil}, P.~C., \& {van Hoof},
  P.~A.~M. 2005, \apj, 624, 794


\bibitem[{{Shaw} {et~al.}(2022){Shaw}, {Ferland}, \&
  {Chatzikos}}]{2022ApJ...934...53S}
{Shaw}, G., {Ferland}, G.~J., \& {Chatzikos}, M. 2022, \apj, 934, 53


\bibitem[{{Shaw} {et~al.}(2020){Shaw}, {Ferland}, \&
  {Ploeckinger}}]{2020RNAAS...4...78S}
{Shaw}, G., {Ferland}, G.~J., \& {Ploeckinger}, S. 2020, Research Notes of the
  American Astronomical Society, 4, 78


\bibitem[{{Si} {et~al.}(2017){Si}, {Li}, {Wang}, {Guo}, {Chen}, {Yan}, {Chen},
  {Brage}, \& {Zou}}]{2017A&A...600A..85S}
{Si}, R., {Li}, S., {Wang}, K., {Guo}, X.~L., {Chen}, Z.~B., {Yan}, J., {Chen},
  C.~Y., {Brage}, T., \& {Zou}, Y.~M. 2017, \aap, 600, A85


\bibitem[{{Smyth} {et~al.}(2019){Smyth}, {Ramsbottom}, {Keenan}, {Ferland}, \&
  {Ballance}}]{2019MNRAS.483..654S}
{Smyth}, R.~T., {Ramsbottom}, C.~A., {Keenan}, F.~P., {Ferland}, G.~J., \&
  {Ballance}, C.~P. 2019, \mnras, 483, 654


\bibitem[{{Tayal} \& {Zatsarinny}(2018)}]{2018PhRvA..98a2706T}
{Tayal}, S.~S. \& {Zatsarinny}, O. 2018, \pra, 98, 012706


\bibitem[{{Temple} {et~al.}(2023){Temple}, {Matthews}, {Hewett}, {Rankine},
  {Richards}, {Banerji}, {Ferland}, {Knigge}, \&
  {Stepney}}]{2023MNRAS.523..646T}
{Temple}, M.~J., {Matthews}, J.~H., {Hewett}, P.~C., {Rankine}, A.~L.,
  {Richards}, G.~T., {Banerji}, M., {Ferland}, G.~J., {Knigge}, C., \&
  {Stepney}, M. 2023, \mnras, 523, 646


\bibitem[{{Tobo{\l}a} {et~al.}(2008){Tobo{\l}a}, {Lique}, {K{\l}os}, \&
  {Cha{\l}asi{\'n}ski}}]{2008JPhB...41o5702T}
{Tobo{\l}a}, R., {Lique}, F., {K{\l}os}, J., \& {Cha{\l}asi{\'n}ski}, G. 2008,
  Journal of Physics B Atomic Molecular Physics, 41, 155702


\bibitem[{{van der Tak} {et~al.}(2020){van der Tak}, {Lique}, {Faure}, {Black},
  \& {van Dishoeck}}]{2020Atoms...8...15V}
{van der Tak}, F. F.~S., {Lique}, F., {Faure}, A., {Black}, J.~H., \& {van
  Dishoeck}, E.~F. 2020, Atoms, 8, 15


\bibitem[{{van Hoof} {et~al.}(2020){van Hoof}, {Van de Steene}, {Guzm{\'a}n},
  {Dehghanian}, {Chatzikos}, \& {Ferland}}]{vanHoof2020}
{van Hoof}, P.~A.~M., {Van de Steene}, G.~C., {Guzm{\'a}n}, F., {Dehghanian},
  M., {Chatzikos}, M., \& {Ferland}, G.~J. 2020, Contributions of the
  Astronomical Observatory Skalnate Pleso, 50, 32


\bibitem[{{van Regemorter}(1962)}]{VanRegemorter1962}
{van Regemorter}, H. 1962, \apj, 136, 906


\bibitem[{{Verner} {et~al.}(1996){Verner}, {Ferland}, {Korista}, \&
  {Yakovlev}}]{1996ApJ...465..487V}
{Verner}, D.~A., {Ferland}, G.~J., {Korista}, K.~T., \& {Yakovlev}, D.~G. 1996,
  \apj, 465, 487


\bibitem[{{Verner} {et~al.}(2000){Verner}, {Verner}, {Baldwin}, {Ferland}, \&
  {Martin}}]{2000ApJ...543..831V}
{Verner}, E.~M., {Verner}, D.~A., {Baldwin}, J.~A., {Ferland}, G.~J., \&
  {Martin}, P.~G. 2000, \apj, 543, 831


\bibitem[{{Verner} {et~al.}(1999){Verner}, {Verner}, {Korista}, {Ferguson},
  {Hamann}, \& {Ferland}}]{1999ApJS..120..101V}
{Verner}, E.~M., {Verner}, D.~A., {Korista}, K.~T., {Ferguson}, J.~W.,
  {Hamann}, F., \& {Ferland}, G.~J. 1999, \apjs, 120, 101


\bibitem[{{Vestergaard} \& {Wilkes}(2001)}]{2001ApJS..134....1V}
{Vestergaard}, M. \& {Wilkes}, B.~J. 2001, \apjs, 134, 1


\bibitem[{{Vincent} {et~al.}(2007){Vincent}, {Spielfiedel}, \&
  {Lique}}]{2007A&A...472.1037V}
{Vincent}, L.~F.~M., {Spielfiedel}, A., \& {Lique}, F. 2007, \aap, 472, 1037


\bibitem[{{Vriens} \& {Smeets}(1980)}]{Vriens1980}
{Vriens}, L. \& {Smeets}, A.~H.~M. 1980, \pra, 22, 940


\bibitem[{{Vrinceanu} \& {Flannery}(2001)}]{Vrinceanu2001}
{Vrinceanu}, D. \& {Flannery}, M.~R. 2001, \pra, 63, 032701


\bibitem[{{Vrinceanu} {et~al.}(2019){Vrinceanu}, {Onofrio}, {Oonk}, {Salas}, \&
  {Sadeghpour}}]{Vrinceanu2019}
{Vrinceanu}, D., {Onofrio}, R., {Oonk}, J.~B.~R., {Salas}, P., \& {Sadeghpour},
  H.~R. 2019, \apj, 879, 115


\bibitem[{{Vrinceanu} {et~al.}(2012){Vrinceanu}, {Onofrio}, \&
  {Sadeghpour}}]{VOS2012}
{Vrinceanu}, D., {Onofrio}, R., \& {Sadeghpour}, H.~R. 2012, \apj, 747, 56


\bibitem[{{Vrinceanu} {et~al.}(2017){Vrinceanu}, {Onofrio}, \&
  {Sadeghpour}}]{Vrinceanu2017}
---. 2017, \mnras, 471, 3051


\bibitem[{{Wang} {et~al.}(2016){Wang}, {Ferland}, {Yang}, {Wang}, \&
  {Zhang}}]{2016ApJ...824..106W}
{Wang}, T., {Ferland}, G.~J., {Yang}, C., {Wang}, H., \& {Zhang}, S. 2016,
  \apj, 824, 106


\bibitem[{{Willacy} \& {Cherchneff}(1998)}]{1998A&A...330..676W}
{Willacy}, K. \& {Cherchneff}, I. 1998, \aap, 330, 676


\bibitem[{{Yan} {et~al.}(2016){Yan}, {Bundy}, {Law}, {Bershady}, {Andrews},
  {Cherinka}, {Diamond-Stanic}, {Drory}, {MacDonald}, {S{\'a}nchez-Gallego},
  {Thomas}, {Wake}, {Weijmans}, {Westfall}, {Zhang}, {Arag{\'o}n-Salamanca},
  {Belfiore}, {Bizyaev}, {Blanc}, {Blanton}, {Brownstein}, {Cappellari},
  {D'Souza}, {Emsellem}, {Fu}, {Gaulme}, {Graham}, {Goddard}, {Gunn},
  {Harding}, {Jones}, {Kinemuchi}, {Li}, {Li}, {Maiolino}, {Mao}, {Maraston},
  {Masters}, {Merrifield}, {Oravetz}, {Pan}, {Parejko}, {Sanchez}, {Schlegel},
  {Simmons}, {Thanjavur}, {Tinker}, {Tremonti}, {van den Bosch}, \&
  {Zheng}}]{yan2016b}
{Yan}, R., {Bundy}, K., {Law}, D.~R., {Bershady}, M.~A., {Andrews}, B.,
  {Cherinka}, B., {Diamond-Stanic}, A.~M., {Drory}, N., {MacDonald}, N.,
  {S{\'a}nchez-Gallego}, J.~R., {Thomas}, D., {Wake}, D.~A., {Weijmans}, A.-M.,
  {Westfall}, K.~B., {Zhang}, K., {Arag{\'o}n-Salamanca}, A., {Belfiore}, F.,
  {Bizyaev}, D., {Blanc}, G.~A., {Blanton}, M.~R., {Brownstein}, J.,
  {Cappellari}, M., {D'Souza}, R., {Emsellem}, E., {Fu}, H., {Gaulme}, P.,
  {Graham}, M.~T., {Goddard}, D., {Gunn}, J.~E., {Harding}, P., {Jones}, A.,
  {Kinemuchi}, K., {Li}, C., {Li}, H., {Maiolino}, R., {Mao}, S., {Maraston},
  C., {Masters}, K., {Merrifield}, M.~R., {Oravetz}, D., {Pan}, K., {Parejko},
  J.~K., {Sanchez}, S.~F., {Schlegel}, D., {Simmons}, A., {Thanjavur}, K.,
  {Tinker}, J., {Tremonti}, C., {van den Bosch}, R., \& {Zheng}, Z. 2016, \aj,
  152, 197


\bibitem[{{Zanchet} {et~al.}(2018){Zanchet}, {Roncero}, {Ag{\'u}ndez}, \&
  {Cernicharo}}]{2018ApJ...862...38Z}
{Zanchet}, A., {Roncero}, O., {Ag{\'u}ndez}, M., \& {Cernicharo}, J. 2018,
  \apj, 862, 38


\bibitem[{{Zhang}(1996)}]{Zhang1996}
{Zhang}, H. 1996, \aaps, 119, 523


\bibitem[{{Zhang} \& {Sampson}(1987)}]{1987ApJS...63..487Z}
{Zhang}, H. \& {Sampson}, D.~H. 1987, The Astrophysical Journal Supplement
  Series, 63, 487


\end{thebibliography}

\end{document}